# Nuclear clocks for testing fundamental physics

E. Peik[1], T. Schumm[2], M. Safronova[3], A. Pálffy[4], J. Weitenberg[5,6] and P.G. Thirolf[7]

[1] Physikalisch-Technische Bundesanstalt, Braunschweig, Germany
[2] Institute for Atomic and Subatomic Physics, Vienna University of Technology, Vienna, Austria
[3] Department of Physics and Astronomy, University of Delaware, Newark, USA
[4] Department of Physics, Friedrich Alexander University Erlangen-Nuremberg, Germany
[5] Chair for Laser Technology, RWTH Aachen University, Aachen, Germany
[6] Max-Planck-Institute of Quantum Optics, Garching, Germany
[7] Department of Medical Physics, Ludwig-Maximilians-University, Munich, Germany

E-mail: xxx@xxx.xx

**Abstract**

The low-energy, long-lived isomer in $^{229}$Th, first studied in the 1970s as an exotic feature in nuclear physics, continues to inspire a multidisciplinary community of physicists. It has stimulated innovative ideas and studies that expand the understanding of atomic and nuclear structure of heavy elements and of the interaction of nuclei with bound electrons and coherent light. Using the nuclear resonance frequency, determined by the strong and electromagnetic interactions inside the nucleus, it is possible to build a highly precise nuclear clock that will be fundamentally different from all other atomic clocks based on resonant frequencies of the electron shell.

The nuclear clock will open opportunities for highly sensitive tests of fundamental principles of physics, particularly in searches for violations of Einstein's equivalence principle and for new particles and interactions beyond the standard model. It has been proposed to use the nuclear clock to search for variations of the electromagnetic and strong coupling constants and for dark matter searches.

The $^{229}$Th nuclear optical clock still represents a major challenge in view of the tremendous gap of nearly 17 orders of magnitude between the present uncertainty in the nuclear transition frequency (about 0.2 eV, corresponding to ~48 THz) and the natural linewidth (in the mHz range). Significant experimental progress has been achieved in recent years, which will be briefly reviewed. Moreover, a research strategy will be outlined to consolidate our present knowledge about essential $^{229m}$Th properties, to determine the nuclear transition frequency with laser spectroscopic precision, realize different types of nuclear clocks and apply them in precision frequency comparisons with optical atomic clocks to test fundamental physics. Two avenues will be discussed: laser-cooled trapped $^{229}$Th ions that allow experiments with complete control on the nucleus-electron interaction and minimal systematic frequency shifts, and Th-doped solids enabling experiments at high particle number and in different electronic environments.

**Table of contents**

## 1  Introduction

The atomic clock is a prominent example of quantum technology that is now sometimes referred to as of a "first generation" [1]. It was devised by I. Rabi and his group in the early 1940s and has been progressively introduced in all fields of precision metrology of time and frequency since the mid-1950s [2]. It has enabled the realization of global navigation satellite systems like GPS, a technology that is now applied ubiquitously in mass-market devices. The development of atomic clocks has pushed the frontier of accuracy of time and frequency measurements since the 1950s by about ten orders of magnitude into the 10$^{-18}$ range [3], much further than for any other physical quantity. At the limits of the highest precision one is usually more concerned with frequency comparisons than with time interval measurements, and one should speak about frequency standards rather than about clocks, but the latter term is widely used in the interest of brevity.

Fig. 1.1 shows the generic schematic of the main building blocks of an atomic clock. The reference of the clock is a selected transition frequency $\nu$ associated with the electromagnetic radiation that is absorbed in transitions between two atomic energy levels separated in energy by $\Delta E$ according to $\nu = \Delta E/h$. For suitably chosen energy levels, the resonance line width can be below 1 Hz, while line shifts due to external perturbations can even be controlled at the mHz level. Consequently, for an optical frequency in the range of $10^{15}$ Hz, the fractional uncertainty range of $10^{-19}$ becomes accessible [4].

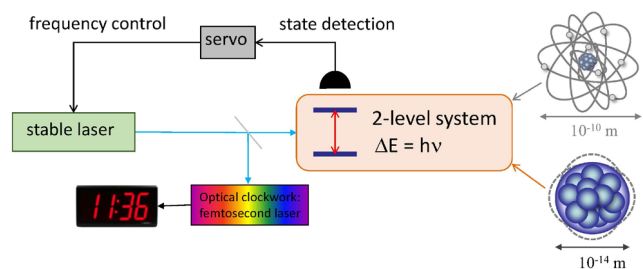

**Fig. 1.1:** Schematic of an optical atomic clock. The laser oscillator frequency is stabilized to the atomic or nuclear resonance frequency and an optical clockwork is employed to produce a time signal.

The oscillator of the clock is a pre-stabilized quartz or other dielectric crystal for the microwave range, or a laser in the case of an optical clock. Clock operation is based on a repetitive cycle, where atoms are first prepared in one of the two energy levels, interrogated with radiation from the oscillator for a certain interaction time, and finally the outcome is measured as the atomic population of the other state. This signal, averaged over several interrogation cycles with the oscillator tuned slightly below and above the center of the resonance curve, is used for the long-term stabilization of the oscillator frequency on the atomic transition frequency. Finally, a clockwork is needed that transfers the periods of the atomic frequency to a standard output signal, like 1 pulse per second or a technically convenient reference frequency like 100 MHz.

Much of the progress with atomic clocks over the past 30 years [3] has been achieved from the application of the methods of laser cooling and trapping of atoms and ions. Stored at sub-mK temperature in ultrahigh vacuum, the localised atoms permit the use of long interrogation times and the low kinetic energy allows for the precise control of the relativistic Doppler shift (or time dilation) from the motion of the atom. A variety of atoms and ions has been investigated as candidates for optical clocks. In many cases, the dominant contributions to the uncertainty budget resulted from the frequency shifts produced by external electric or magnetic fields. This includes some contributions that are difficult to control like electric field gradients in ion traps or the fields of blackbody radiation emitted from surfaces surrounding the atoms. This has provided the background and motivation for the proposal of a nuclear optical clock [5].

The conceptual shift from an atomic to a nuclear clock is straightforward: instead of a transition in the electron shell, a radiative transition inside the nucleus shall be used as the reference for frequency stabilization of a laser oscillator [6]. Since the nucleus itself is much smaller and much less polarizable than the atom, the effect of external electric fields and field gradients can be expected to be much smaller. Two further aspects contributed to the interest in the idea: first, Mössbauer spectroscopy has shown that extremely high resolution may be obtained for nuclear resonances in solids. This would open an opportunity to interrogate possibly $10^{14}$ nuclei instead of the typically $10^4$ laser-cooled atoms in an optical clock, potentially providing a much stronger signal. Second, in contrast to the transition frequency of a valence electron, the nuclear transition frequency is not purely determined by electromagnetic interactions but also by the strong interaction. This makes the nuclear clock an interesting contributor in tests of fundamental physics based on precision clock comparisons, especially for searches for variation of fundamental constants [5,7].

Such considerations would remain remote from practical implementation if we would have to consider only the established Mössbauer transitions with photon energies of tens of keV, because the required frequency-stable laser source and a clockwork for counting of oscillation periods are not available at those frequencies. Fortunately, one suitable nucleus is known for the development of a nuclear clock in the frequency range that is accessible with present laser technology: the low-energy transition in $^{229}$Th. This system has provided the motivation for the development of concepts and experiments that we will review here.

Amongst the presently known about 3300 nuclides with their in total about 184000 nuclear excited states [8], the thorium isomer $^{229m}$Th assumes a unique position, since it represents the by far lowest nuclear excitation in the whole landscape of atomic nuclei, with an energy $E$ in the range of outer-shell electronic transitions and realistically accessible with existing laser technology. Moreover, according to Fermi's Golden Rule, the lifetime $\tau$ of this state scales as $E^{-3}$ for an M1 ground-state transition, such that it is expected to be of the order of a few $10^3$ seconds. In turn, according to $\tau = \hbar/\Delta E$ this results in an extremely narrow relative linewidth of $\Delta E/E \approx 10^{-20}$ for the ground-state transition. Together with a high resilience against external perturbations due to the small nuclear moments, these properties render $^{229m}$Th an attractive (and presently the only) candidate for constructing a nuclear frequency standard that could rival todays most advanced optical atomic clocks. Fig. 1.2 impressively illustrates the uniqueness of the thorium isomer in a display of the half-lives of all known nuclear isomers versus their excitation energies. $^{229m}$Th resides far-off all other isomers surrounded by typical atomic transitions being used in optical atomic clocks. While hints to a potential application of the thorium isomer in the field of frequency metrology can be found already in the early literature on $^{229m}$Th, it was in the pioneering publication by Peik and Tamm in 2003 [5] that for the first time a viable concept was presented for building a 'nuclear clock' based on the thorium isomer $^{229m}$Th.

On a historical note, what is nowadays termed the 'thorium isomer' $^{229m}$Th entered the scene of nuclear physics already in 1976, when $\gamma$-spectroscopic nuclear structure studies on $^{229}$Th could only be consistently interpreted by introducing an almost degenerate ground-state doublet in $^{229}$Th, following the $\alpha$-decay of $^{233}$U [9]. Initially, an excitation energy of less than 100 eV was estimated, since no direct evidence for the ground-state decay of this new first excited state could be resolved. Over the following decades, many experimental attempts were conducted towards further constraining the excitation energy of this exceptionally low-lying nuclear excited state [10–18]. These early efforts converged in 1994 to an energy value of 3.5(10) eV [12]. Focusing on the excitation energy, interestingly, the isomeric nature of this state was not explicitly mentioned or discussed during the first almost two decades. This aspect was first addressed in 1994 when the half-life was conjectured from the (hindered) pure M1 characteristics of the ground-state transition to lie in the range of 20-120 hours [12]. Already in this publication it was speculated that for $^{229m}$Th no unique half-life might exist, but rather the coupling to and the exact nature of the electronic environment had to be taken into account. Following this, for more than a decade the name '3.5 eV isomer' was used in the literature. From today's perspective the experimental efforts at the time were hampered by the fact that assuming an excitation energy around 3.5 eV, corresponding to wavelengths in the optical to near-UV range, radiative decay was considered the only possible decay branch, as the higher-lying ionization potential of thorium (today known as 6.308(3) eV [19]) would energetically not allow for internal-conversion (IC) decay. All related experimental efforts remained inconclusive [20–28], as detailed later in Sec. 2.2. Experimental studies were accompanied by theoretical efforts to better understand the specific properties of the thorium isomer [29–31] or propose various ways for its population [30,32–37]. See also the review articles [38,39].

The situation changed when a measurement using a novel X-ray microcalorimeter in 2007 raised the excitation energy to 7.6(5) eV [40], later re-analyzed and corrected to 7.8(5) eV ($\lambda = 160(10)$ nm) [41]. For the first time closely-spaced doublets of rotational $\gamma$-ray transitions around 29 keV and 42 keV, respectively, could be resolved, allowing to indirectly infer the isomer's excitation energy. This shifted the transition energy into the vacuum-ultraviolet (VUV) range and above the first ionization potential of thorium. The new result triggered numerous efforts towards direct observation and energy determination of the nuclear transition. Despite the considerably raised value of the excitation energy, the most commonly used methods were (VUV) fluorescence detection from $^{229}$Th-doped crystals and optical excitation in trapped $^{229}$Th ions [42–50]. These efforts went along with proposals of various excitation and de-excitation schemes [51–54], preparations for laser-spectroscopic studies [55,56] and quantitative assessments of the available parameter space for half-life and excitation energy [57]. It was only in the most recent years that a change in paradigm allowed the first direct observation of the isomeric decay in the process of internal conversion, and coined more precise values for the isomer energies. In the following section our current knowledge of the properties of the thorium isomer and its ground-state decay will be presented as the basis for all ongoing and envisaged activities towards building a nuclear clock and applying it for fundamental physics tests. Further details on historical aspects, experimental procedures and application potentials of the thorium isomer, not discussed here in depth, can be found in the review articles [58–60].

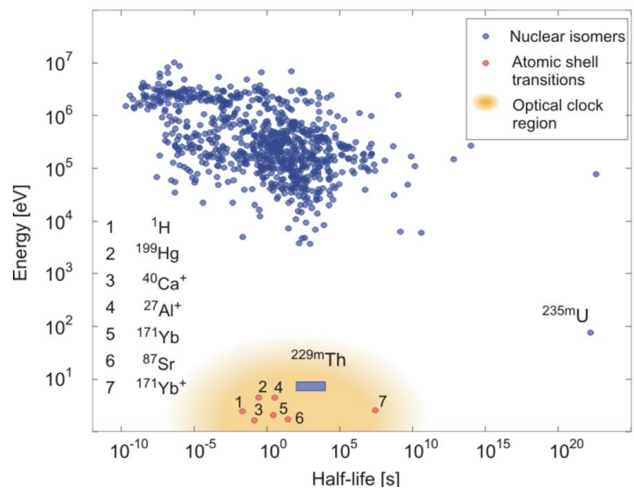

**Fig. 1.2:** Distribution of excitation energy versus half-life of isomeric nuclear states. Nuclear excitations (blue circles) exhibit typical energies in the range of a few 10 keV up to several MeV. Only two low-energy (<1 keV) nuclear states are presently known: $^{229m}$Th (∼8.2 eV, expected energy range indicated by the blue box) and $^{235m}$U (76.7 eV). Due to the extremely small transition strength in $^{235m}$U (radiative lifetime $\sim 10^{22}$ s), only $^{229m}$Th qualifies for a direct laser excitation and thus for the development of a nuclear clock. In addition, selected atomic shell transitions are included (red circles), which are already in use for optical atomic clocks [61].

Following this Introduction, Sec. 2 presents our current knowledge on the properties of the $^{229m}$Th nuclear clock transition. Sec. 3 sketches the strategy towards the realization of a nuclear clock along different experimental avenues and supported by theoretical guidance. Sec. 4 outlines the potential of the nuclear clock for testing fundamental physics. The paper focuses on the planned efforts of the teams forming the 'ThoriumNuclearClock' consortium within a 'Synergy' project framework of the European Research Council (ERC) [62].

## 2 Present knowledge of the $^{229m}$Th nuclear clock transition

Our present knowledge of key properties for ground and isomeric excited states as well as for the nuclear clock ground-state transition from $^{229m}$Th is summarized in Fig. 2.1. Nuclear spins, parities and Nilsson quantum numbers of both orbitals are displayed together with their magnetic dipole and electric quadrupole moments. The properties of the ground-state transition from $^{229m}$Th are highlighted in bold font.

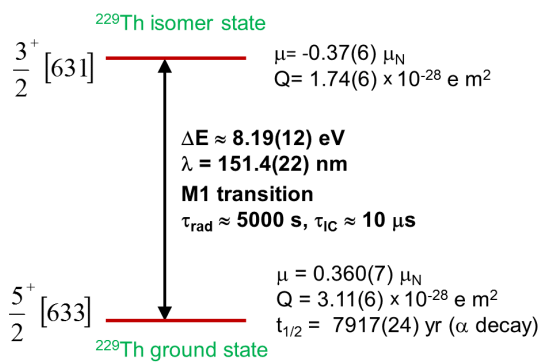

**Fig. 2.1:** Properties of the $^{229}$Th ground-state doublet: Nuclear levels are given with their spin, parity and dominant Nilsson classification. $\mu$: magnetic moment in nuclear magnetons $\mu_N$, $Q$: spectroscopic electric quadrupole moment. The isomer's excitation energy and wavelength is given as the average between the values from [66] and the most precise value from [67]. Uncorrelated statistical uncertainties were assumed for the determination of the combined uncertainty.

### 2.1 Studies based on internal conversion

Many of these properties have been recently determined or refined, following the pioneering experiment in 2016 that succeeded in the realization of the first direct detection of the thorium isomer's ground-state decay, not in the radiative decay channel but via the detection of electrons emitted in the IC decay of $^{229m}$Th [61]. In the process of IC, the nuclear excitation energy is transferred to the atomic shell, resulting in the ejection of an atomic shell electron. The key to the success of this experiment is shown in Fig. 2.2. A buffer-gas stopping cell is the main component of the experimental setup [63] that allows to generate an isotopically clean beam of $^{229}$Th ions, starting from the $\alpha$-decay of $^{233}$U [64,65]. Fortunately, 2% of the $\alpha$-decay do not directly populate the ground state in $^{229}$Th, but pass via the isomeric first excited state. Until today this represents the traditional population pathway of the thorium isomer.

Decoupling the population of the isomeric state in the gas cell from its delayed decay about a meter away enables to discriminate all prompt background generated during the $\alpha$-decay. Inside the gas cell, starting from the solid $^{233}$U source, the recoiling $\alpha$-decay daughter nuclei with kinetic energies of about 84 keV are thermalized within 1-2 cm in an ultra-pure He buffer gas atmosphere of about 30 mbar. Ultimate (UHV-grade, including baking to 130°C) cleanliness of the gas cell allows for an efficient extraction of recoil ions without significant losses by, e.g., charge exchange or formation of molecules. Being emitted isotropically into the gas volume, a conical RF+DC funnel system is employed to guide the stopped recoil ions towards the exit from the gas cell, which is formed by a supersonic de-Laval-type nozzle, electrically isolated in order to serve as last extraction electrode. In the region of the 0.6 mm diameter nozzle a supersonic gas jet is formed that rips the ions off the electrical field lines and drags them with the gas jet into the subsequent vacuum chamber that houses a 12-fold segmented radiofrequency quadrupole (RFQ) ion guide, buncher and phase-space cooler. Accompanying $\alpha$-decay daughter products from the $^{233}$U decay chain are finally filtered out by a quadrupole mass separator (QMS). More details about this 'thorium isomer generator', able to provide clean $^{229(m)}$Th beams as singly to triply charged ions, can be found in [64]. Behind the QMS (and a triodic extraction electrode system) a microchannel plate (MCP) detector allows for the registration of the decay products emitted in the $^{229}$Th isomeric decay. The diagnostics behind the QMS was laid out for the detection of conversion electrons on the MCP detector, based on a twofold reasoning:

(i) The revised excitation energy of the thorium isomer from the Beck et al. experiment [38] had opened the IC decay channel for neutral $^{229m}$Th. (ii) Theoretical calculations [23,51] expected such a decay to strongly dominate any competing radiative decay by a huge conversion factor of $\alpha_{IC} = \Gamma_e/\Gamma_\gamma \approx 10^9$ (with $\Gamma_e$ and $\Gamma_\gamma$ being the decay rates for IC and $\gamma$-decay, respectively).

Extracted $^{229m}$Th ions directly impinge on the MCP surface, set to a mildly attractive potential to secure 'soft landing' of the ions without generating a signal from secondary electrons in the MCP detector. Thorium ions would immediately neutralize and thus trigger IC decay, emitting conversion electrons that will be amplified, electrostatically accelerated behind the MCP towards a phosphor screen and optically registered by a CCD camera.

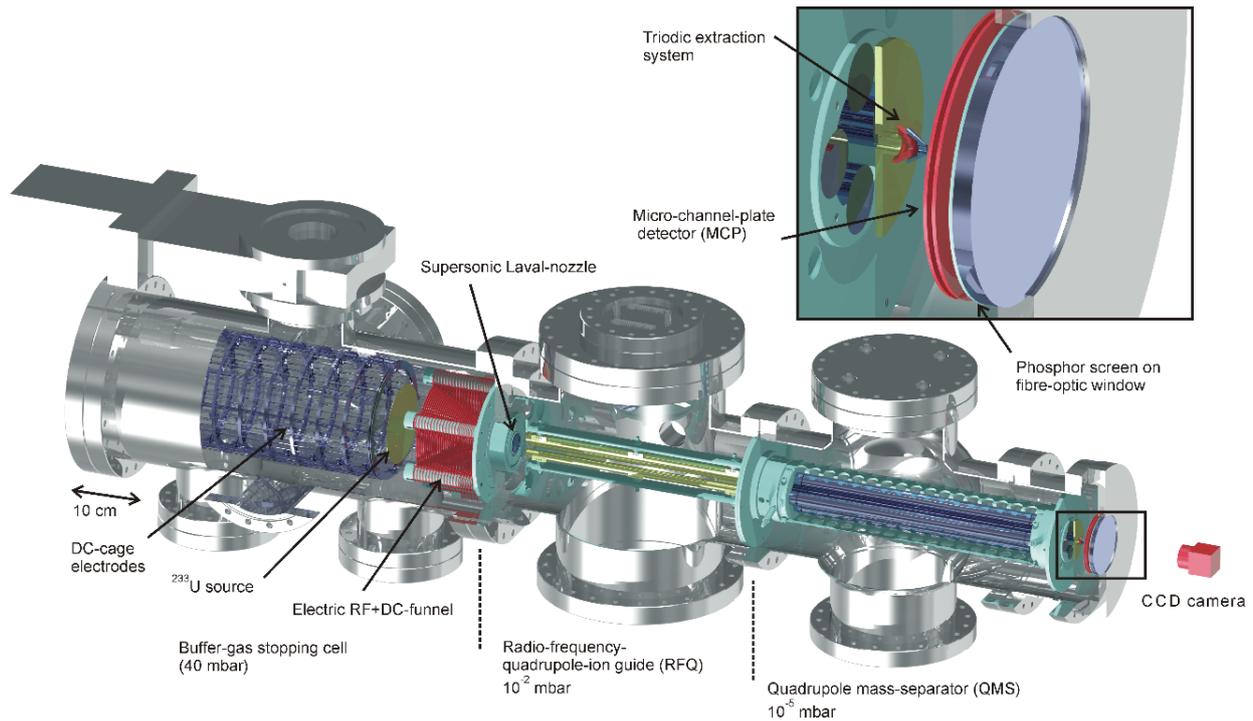

**Fig. 2.2:** 3D technical drawing of the experimental setup employed for the generation of an isotopically clean $^{229(m)}$Th ion beam. It consists (from left to right) of a buffer-gas stopping cell that houses a $^{233}$U $\alpha$ recoil source, a (segmented) radio-frequency quadrupole (RFQ) as ion guide, buncher and phase space cooler, a quadrupole mass separator (QMS) and (behind a triodic extraction electrode system) a multichannel-plate detector followed by a phosphor screen and a CCD camera [61]. For more details see [59].

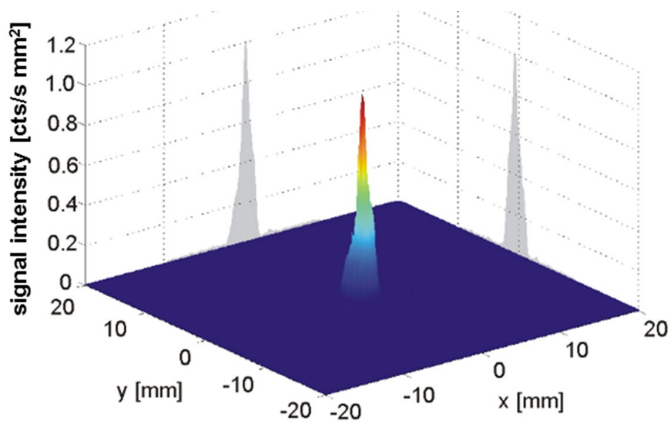

**Fig. 2.3:** Signal of the first direct detection of the $^{229m}$Th decay. Shown is the IC electron signal of $^{229m}$Th$^{3+}$ ions, which were collected with low kinetic energy directly on the surface of a position-sensitive MCP detector [61].

This experimental arrangement led to the first direct detection of the thorium isomer's ground state (IC) decay [61]. Fig. 2.3 shows this breakthrough result of the $^{229m}$Th IC electron signal obtained with a high signal-to-background ratio, which subsequently triggered renewed and intensified experimental and theoretical efforts towards the realization of a $^{229m}$Th-based nuclear clock.

Only little adaptation of the above-described experimental setup and procedure was needed to achieve a first lifetime measurement of the thorium isomer under IC decay. An illustration of the setup (panel b) and the detection scheme is presented in Fig. 2.4. The segmented extraction RFQ was operated as linear Paul trap and served to extract sharp ion bunches from the buffer gas cell before sending them to the MCP detector.

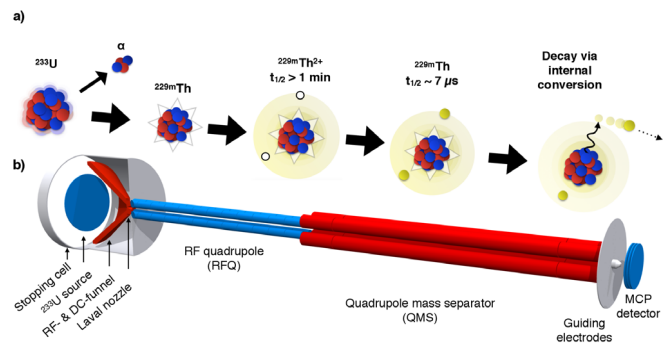

**Fig. 2.4:** a) Visualization of the detection scheme reported in [68]. An ion beam is formed from $^{229(m)}$Th ions emerging from a $^{233}$U $\alpha$-recoil source. As long as $^{229m}$Th remains in a charged state, the lifetime is longer than 1 minute. The moment the ions neutralize (e.g. by collecting the ions on a metal surface) internal conversion is triggered and the lifetime reduced to the range of μs. The emitted electron can be detected. b) Scheme of the experimental setup that was used [68].

However, measuring the thorium isomer's radiative lifetime in ionic charge state was not yet possible for the expected few $10^3$ s, due to limitations imposed by the achievable high-vacuum conditions in the Paul trap, where ions could not be stored for longer than about 1 minute. Therefore, the measurement addressed instead the lifetime of the $^{229m}$Th internal-conversion decay. Again, a $^{229(m)}$Th ion beam was formed within several milliseconds, which was then accumulated directly on the metallic surface of the MCP detector which was used both the charge capture of the ions and the detection of the IC electron. As soon as the $^{229(m)}$Th ions neutralize, IC is initiated and an electron is emitted and registered with the MCP detector. Fig. 2.5 displays the resulting electron time distributions, where $^{230}$Th$^{3+}$ exhibits only the strong ionic impact signal during the pulse duration of about 10 µs, while for $^{229}$Th$^{3+}$ the exponential decay tail of the isomer is clearly visible.

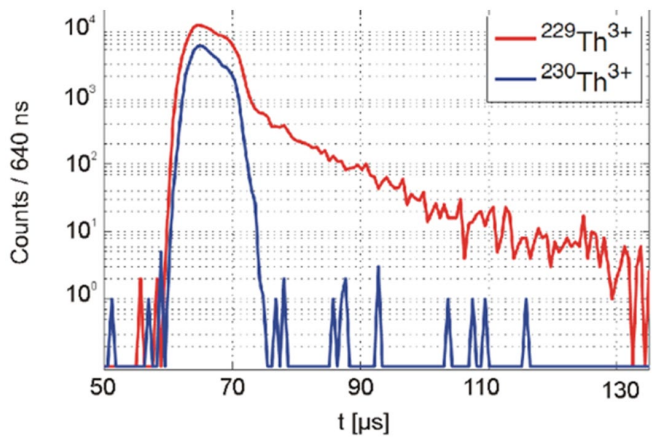

**Fig. 2.5:** Measurement of the $^{229m}$Th IC lifetime in neutral thorium. Pulsed $^{229}$Th$^{3+}$ ions impinge onto a metallic surface, where they neutralize, followed by time-resolved detection of the resulting IC electrons (red). The measure half-life amounts to about 7 µs. In addition, the identical measurement is shown for $^{230}$Th$^{3+}$ ions (blue), where only the signal of the ionic impact is observed [68].

The determined short half-life of about 7(1) µs nicely agrees with the theoretically expected drastic lifetime reduction of the thorium isomer in case of IC decay by about 9 orders of magnitude [31,57]. Nevertheless, a word of caution has to be issued, as the lifetime of the isomer will depend on the electronic environment of the nucleus and thus on the chemical structure of the specific surface used to neutralize the ion and hence induce IC decay (see Sec. 3.1.2.1 for further discussion). Clear effects of such dependencies have already been observed during the measurements of [68], reported in[69], while systematic studies using a variety of metallic surfaces will be performed in the near future. Most notably, no IC signal could be found for singly charged $^{229m}$Th ions. Although extracted from the gas cell much less abundant than doubly and triply charged ions, a clear signal should have been observed as well. Its non-observation is a strong hint towards a short lifetime (significantly below the ion extraction time of a few ms [63]) of the isomeric state in singly charged $^{229}$Th. Possible explanations could be quenching of the isomer in the buffer gas or a bound internal-conversion decay of the isomer in the 1+ charge state [70].

The opportunity to trap Th ions in an ensemble containing nuclear ground states and isomers has enabled a laser spectroscopic investigation of hyperfine structure that has provided the first experimental values of nuclear moments of the isomer and its *rms* charge radius. The experiment has been performed with Th$^{2+}$ because of its higher abundance from the recoil ion source and longer trap lifetime in comparison to Th$^+$. Since the trapped ions exchange kinetic energy with the buffer gas, they are approximately at room temperature and Doppler broadening to about 1 GHz linewidth impedes the complete resolution of the hyperfine structure. Therefore, a two-step laser excitation scheme has been employed, where the first laser selectively excites one narrow velocity class out of the thermal ensemble and a second laser probes the hyperfine structure on these ions, free from Doppler broadening. Varying the frequencies of both lasers, the resolved hyperfine structure (see Fig. 2.6) could be recorded for both electronic transitions, including the isomer shift between $^{229m}$Th$^{2+}$ and $^{229}$Th$^{2+}$. In comparison to reference spectra recorded using the same method with $^{229}$Th$^{2+}$ and $^{232}$Th$^{2+}$, produced in laser ablation from a solid sample, it has been possible to unambiguously identify the hyperfine components of the isomer, marking its first non-destructive optical detection.

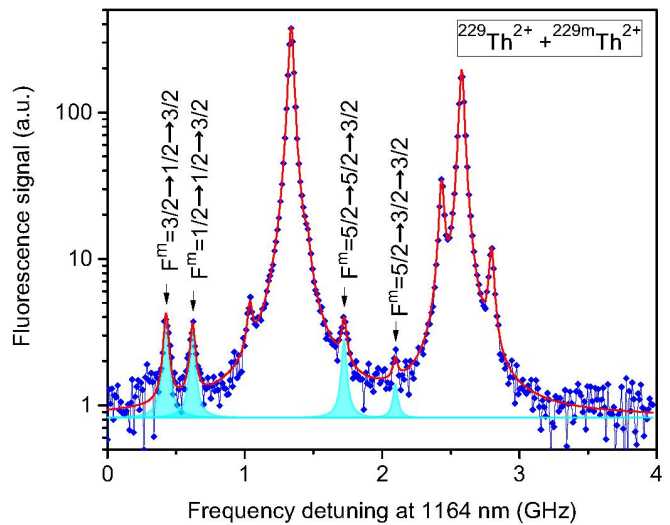

**Fig. 2.6:** Hyperfine structure resonances of nuclear isomeric and ground states. The spectrum results from a two-step laser excitation that eliminates Doppler broadening, showing the relative strengths (in logarithmic scale) and frequency range of nuclear isomeric and ground state resonances. Resonances of the nuclear isomer HFS are marked in cyan color. The unlabeled peaks belong to the nuclear ground state [71].

The magnetic dipole moment of the isomer $\mu^m = -0.37(6)\,\mu_N$ turned out to be 5 times larger than a prediction based on the Nilsson model [34]. This finding has stimulated a refinement of the $^{229}$Th nuclear models [72]. The intrinsic electric quadrupole moment of the isomer $Q_0^m = 8.7(3)\,eb$ is identical to that of the ground state within the uncertainty, indicating that the shape and size of the proton distribution changes only little in the nuclear excitation from the ground state to the isomer. This is also indicated by the very small difference in *rms* charge radius $\langle r^2 \rangle^{229m} - \langle r^2 \rangle^{229} = 0.0105(13)$ fm$^2$ [73] that has been determined from the comparison of isomer and isotope shifts relative to the radius difference between $^{229}$Th and $^{232}$Th. Apart from providing first information on the internal properties of the isomer, these numbers are also of relevance for estimating the sensitivity of a $^{229}$Th nuclear clock in fundamental tests (see Sec. 4).

Coming back to the quest of more precisely determining the isomer energy, the 9 orders of magnitude stronger IC decay obviously presents some advantages compared to the radiative VUV fluorescence decay channel. Consequently, a direct measurement of the ground-state transition energy of this isomeric state was performed based on the spectroscopy of the internal-conversion electrons emitted in flight during the decay of neutral $^{229m}$Th atoms.

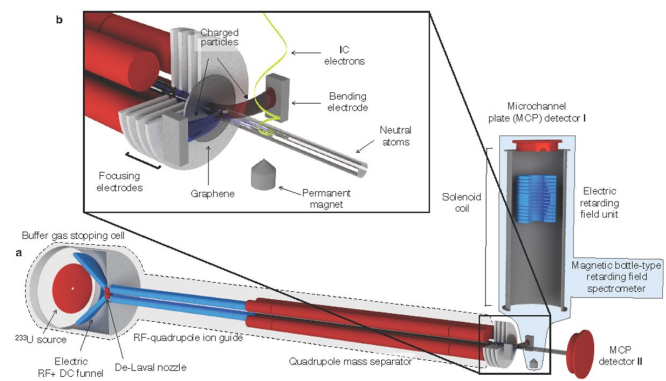

**Fig. 2.7:** Experimental setup used for the direct measurement of the $^{229m}$Th excitation energy via a determination of the kinetic energy of IC electrons in a magnetic-bottle-type electron spectrometer. The front part is identical to the setup to generate a clean $^{229(m)}$Th ion beam, already described in Fig. 2.2 and Fig. 2.4, now complemented by the electron spectrometer (right-hand side) placed behind a graphene neutralization foil and an ion removal electrode (also upper left inset) [66].

Since the setup (Figs. 2.2 and 2.4) did not allow for measuring the kinetic energy of IC electrons, it was extended by an ion-neutralizer (shown in the Fig. 2.7b) and a magnetic bottle-type retarding field electron spectrometer (Fig. 2.7 and Fig. 2.8): The ions are guided by four focusing electrodes onto two layers of graphene set to −300 V. In passing these foils, the ions are neutralized and continue their flight as neutral atoms (Fig. 2.7b). The extraction and neutralization is monitored with a MCP detector placed in the central beam axis (MCP detector II in Fig. 2.7a). The isomer decays within microseconds by emitting an electron. The electron's kinetic energy is measured free from any surface influences with a magnetic bottle-type retarding field electron spectrometer [74], which is placed 90° off-axis behind the graphene (see Fig. 2.7). Bending electrodes (to which a DC field is applied) are placed between the graphene layer and the spectrometer entrance in order to prevent charged particles from entering the spectrometer. IC electrons, which are emitted above a strong permanent magnet, are collected and guided towards a retarding field unit placed in a solenoid coil. The electrons' kinetic energy can be analysed by applying a retarding voltage to a grid and counting the electrons that reach the MCP detector I (see Fig. 2.7b). Fig. 2.8 shows a sectional view of the magnetic-bottle spectrometer and its retarding field unit.

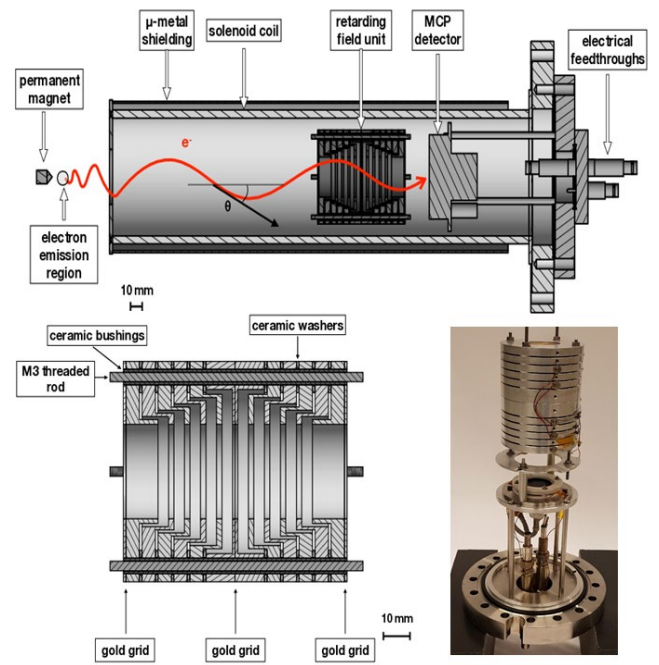

**Fig. 2.8:** Sectional view and photograph of the magnetic-bottle type retardation electron spectrometer. The upper panel shows an overview of the spectrometer, with the permanent magnet, the solenoid coil, the retarding field unit and the MCP detector. The region of electron emission as well as a helical electron trajectory with its pitch angle θ is indicated. The bottom panel (left) shows a detailed view of the retarding field unit. The positions of the gold grids are indicated by arrows. The bottom right panel shows a photograph of the spectrometer. Reprinted (in parts) from [74], Copyright (2019), with permission from Elsevier.

The energy resolution of the spectrometer was determined with argon and neon as calibrant gases and found to be between 2.2% (Ar) and 3% (Ne), corresponding to an energy resolution of 30-50 meV for the IC electron kinetic energies to be expected from the decay of $^{229m}$Th [63]. This efficient spectrometer allowed for the first direct measurement of the thorium isomer's excitation energy. As IC after resonant neutralization occurs from excited electronic states, the kinetic energy $E_{kin}$ of an IC electron is connected to the energy $E_I$ of the isomer via $E_{kin} = E_I - IP + E_i - E_f$, with the thorium ionization potential $IP$ (6.308(3) eV [64]), $E_i$ being the excitation energy of the Th atom undergoing IC decay and $E_f$ as the energy of the final electronic state of the Th ion generated during the IC process. A statistical analysis method was developed in [65], which, together with atomic theory input (density functional calculations performed to understand which electron orbitals become resonant with the valence band of graphene during resonant neutralization and atomic structure and IC calculations to take into account all possible excited initial and corresponding final electronic states and related IC rates), allowed for deriving a value for the isomeric excitation energy, for the first time directly measured and with improved precision compared to the long-time adopted value: $E_I = 8.28(3)_{stat.}(16)_{syst.}$ eV = 8.28(17) eV [66]. This energy corresponds to a wavelength of 149.7(31) nm.

## 2.2 Gamma spectroscopy measurements

An alternative approach to indirectly determine the isomer energy is high-precision gamma spectroscopy of the higher-lying excited states of the $^{229}$Th nuclear structure (see Fig. 2.9). This nuclear physics method is tightly intertwined within the $^{229m}$Th discovery quest. When Kroger and Reich measured the gamma spectrum emerging from the $^{233}$U→$^{229}$Th $\alpha$-decay, they noticed that several gamma lines connected with the 3/2 [631] excited state were missing [9]. This could only be explained by a near-degeneracy of the 3/2 [631] isomeric state with the 5/2 [633] ground state. The experimental resolution of 450 eV allowed to place an upper bound of $E_I < 100$ eV on the isomer energy, making it the lowest-energy excited state of the $^{229}$Th nucleus by far. A similar experiment was repeated in 1985, predicting an isomer energy of −1(4) eV [75].

In 1994, a further improved version of the experiment was performed, yielding a value of $E_I = 3.5(10)$ eV, placing it in the visible part of the electromagnetic spectrum, around 355 nm wavelength. This value remained the "accepted" value for more than a decade. As already mentioned in the Introduction, a series of attempts were made to detect the "optical" gamma ray that should emerge when the isomer state (populated with 2% probability in $^{233}$U $\alpha$-decay) decays to the ground state. Several publications claimed having observed this signature [20,21], but all observations could later be explained by parasitic effects related to the radioactivity of $^{233}$U [17,22]. A re-analysis of the 1994 data in 2005 including recoil effects predicted a shifted value of 5.5(10) eV, moving the excitation wavelength into the ultraviolet range around 225 nm [23].

In 2007, a new cryogenic microcalorimeter originally developed by NASA for X-ray spectroscopy was used to again record the gamma spectrum emerging in the $^{233}$U→$^{229}$Th $\alpha$-decay, now focussing on lower energy X-ray and gamma lines between 10-50 keV [40]. The spectrometer had a resolution of 30 eV and could resolve the spectrally neighbouring lines at 29.19 keV and 29.39 keV as well as 42.43 keV and 42.63 keV (see Fig. 2.9).

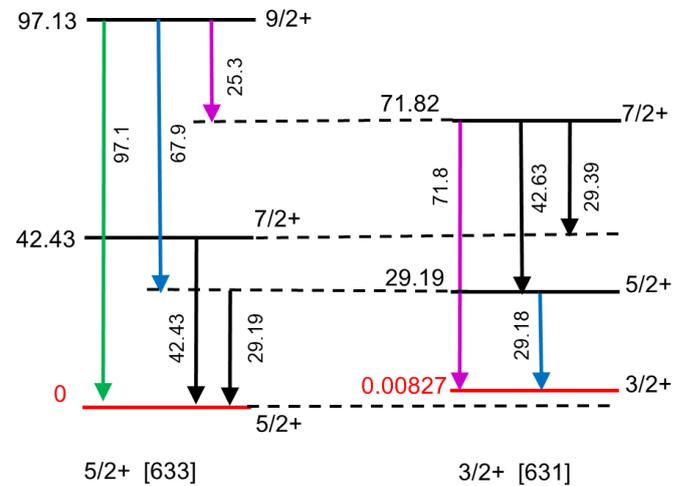

**Fig. 2.9:** Low-energy part of the $^{229}$Th level scheme (energy units in keV). In square brackets the Nilsson quantum number categorization of the band-head levels is indicated.

Measuring these lines allowed to significantly reduce the calibration uncertainty in the data analysis and yielded an isomer energy of 7.6(5) eV. Later, in 2009, this value was shifted to 7.8(5) eV, accounting for unresolved line doublets, assuming theoretically predicted branching ratios [41]. However, errors due to the detector calibration were not considered in the analyses and the overall energy uncertainty is probably higher [76]. The 7.8(5) eV value for the isomer energy was the standing assumption for another decade. The excitation wavelength would now be around 160 nm, in the VUV range. As this wavelength is absorbed by practically all media, it might explain the failure to detect the direct radiative decay by optical means (mostly performed in air) in previous experiments.

In 2020, a new variant of a cryogenic microcalorimeter, specifically developed for determining the $^{229}$Th isomer energy, was used [67]. The maXs30 magnetic microcalorimeter measured the heat deposited in an absorber pixel upon absorption of a gamma or X-ray. This heat deposit changed the magnetization of a Ag:Er$^{3+}$ paramagnetic sensor, the magnetization change was read out by a SQUID magnetometer. The device has a resolution of 10 eV and exceptional gain linearity, reducing the calibration error. The $^{233}$U→$^{229}$Th gamma spectrum was measured in the range between 10 keV and 60 keV for several months, and the data is openly accessible [77]. Several different decay paths could be identified within the $^{229}$Th nuclear structure that allow deducing the isomer energy. The most precise method (also used in [40]) yielded a value of $E_I = 8.10(17)$ eV, consistent with the value derived from IC electron spectroscopy. We note that in this analysis, the isomer energy uncertainty is affected by the overall calibration uncertainty of the microcalorimeter. If new calibration values in the range 10-60 keV with higher accuracy would become available, the precision on the isomer energy could be further improved using the existing spectrometer data.

Beyond the isomer energy, several other relevant nuclear properties could be extracted from this data. First, the line corresponding to the 2$^{nd}$ excited nuclear state at 29.19 keV shows a clear bi-modal structure, which is the doublet that corresponds to decays into the 5/2 ground state or the 3/2 isomer state respectively. From the relative amplitudes of these two doublet lines, the corresponding branching ratio $b_{29}$ could be extracted to 1/10.8 (9.3(6)% into the cross-branch decay to the ground state). The branching ratio of the 3$^{rd}$ excited state $b_{42}$ could be constrained to <0.7%. Again, if branching ratios would be measured with higher accuracy by other methods, the precision on the isomer energy could be increased using the existing gamma spectroscopy data.

It is interesting to note that gamma spectroscopy data from different measurements (different campaigns or different types of measurements) can be combined to extract the $^{229}$Th isomer energy or other nuclear parameters. For example, a gamma measurement of the 29.19 keV line (doublet) has been published in 2019, using a transition edge sensor [78]. As the 29.19 keV level very predominantly decays intra-band, its gamma signature can be combined with an excitation measurement of the same state (see 2.3 below), which exclusively drives the inter-band transition. With a spectral resolution of 40 eV this combined analysis yielded an isomer energy of 8.30(92) eV, which does not reduce the uncertainty but provides an interesting consistency check using a different detector technology. A similar analysis using the before mentioned maXs30 magnetic microcalorimeter data yields $E_I$ = 7.8(8) eV.

Such inter-device-analyses are burdened with the requirement of an extremely accurate energy calibration (relative precision >10$^5$). Currently, there is a need for improved absolute energy measurements of calibration lines that could reduce the calibration error in the already existing datasets. In general, data will have to be iteratively re-evaluated once new, improved measurements become available, highlighting the need for the open-access publishing of raw spectroscopy data.

## 2.3 X-ray pumping, nuclear resonant scattering

Resonant excitation of nuclear states using synchrotron X-ray radiation is an exciting alternative to population through (stochastic) nuclear decay process. Excitation from the 5/2$^+$ [633] ground state to the 5/2$^+$ [631] 2$^{nd}$ excited state at 29.19 keV (inter-band excitation, see Fig. 2.9) has recently been demonstrated using the SPring-8 BL19LXU beamline [79]. In this experiment, a tuneable narrow-band (FWHM linewidth 0.1 eV), high-intensity (10$^{12}$ photons/s), collimated (0.15 × 0.065 mm²) X-ray beam was directed at a solid-state $^{229}$Th sample (1.8 kBq). To identify the excitation energy, the technique of nuclear resonant scattering was used: off-resonance, the synchrotron X-ray beam interacts exclusively with the $^{229}$Th electron shell. As the synchrotron source is pulsed (42 MHz pulse frequency), this photoelectric interaction produces a "prompt" electron scattering signal: electron vacancies are created in the L-shell and successively filled, emitting characteristic X-ray fluorescence in the energy range 12-18 keV.

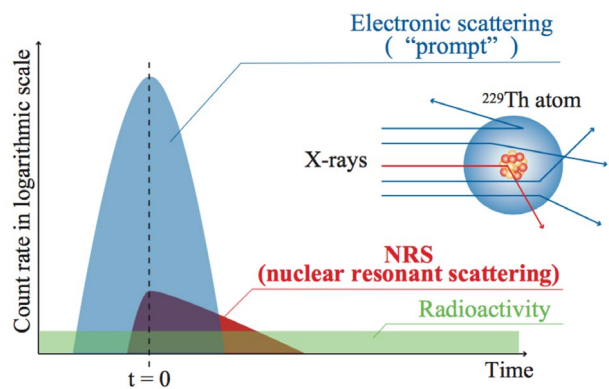

**Fig. 2.10:** Schematic of temporal profiles of the NRS process [79].

On-resonance (with the nuclear excitation energy), this process also takes place, producing a "prompt" background. However, additionally, the nucleus is excited to the 2$^{nd}$ excited state with a certain cross section, referred to as nuclear resonant scattering (NRS) [80]. This nuclear excitation decays again with a characteristic timescale, 82 ps in the case of $^{229}$Th (29.19 keV level). The dominant decay path is internal conversion, where the nuclear energy is transferred to the electron shell. Similar (but not identical) to the photoelectric interaction, this again produces characteristic X-ray lines. The key aspect here is that this NRS signal is delayed by the nuclear excited state lifetime and can hence be distinguished by its timing characteristics (see Fig. 2.10). However, the "prompt" signal supersedes the NRS signal by 6 orders of magnitudes and filtering for a 80 ps delay with such a background is a formidable technological challenge [81].

Scanning the energy of the SPring-8 X-ray beam, the NRS resonance was found and the excitation energy of the 2$^{nd}$ excited state in $^{229}$Th (29.19 keV) was measured to be 29189.93(7) eV. The absolute energy of the X-ray excitation beam was monitored by measuring a pair of Bragg angles on the left/right side of the beam using a Si(440) reference crystal (Bond method) [82,83]. This energy measurement is one of the most precise ever performed in NRS. It is amusing to note that after hunting down the energy of the 1$^{st}$ excited isomeric state of $^{229}$Th for over 40 years, it is the 2$^{nd}$ excited state which is one of the best-known nuclear levels, with a stunning relative precision of 4·10$^6$.

By comparing the signal yields obtained for the photoelectric and the NRS interaction, one can deduce the excitation cross section for the cross-band nuclear excitation to the 29.19 keV state. Combined with the formerly known cross section for the in-band process [9,15], the $b_{29}$ branching ratio could be extracted as 1/9.4(24), in good agreement with the value measured in gamma spectroscopy.

This branching ratio describes the radiative (gamma) decay of the 29.19 keV level, leading predominantly (90%) into the isomeric state (in-band transition). However, as mentioned before, the 29.19 keV state mostly decays via internal conversion, the ratio of IC-decay to gamma-decay is predicted to be around 225 [8]. Combining both radiative and IC decay processes, the probability to reach the isomeric state from the 29.19 keV level is 0.58(7). Combining this value with the excitation rate to the 29.19 keV level provided by the SPring-8 X-ray excitation (on resonance) yields a pumping rate of 25000 s$^{-1}$. NRS excitation to the 2$^{nd}$ excited nuclear state with subsequent decay hence represents a very efficient means to pump population into the isomeric state. Besides the $^{233}$U decay, it is the second proven isomer population method today. It has the great advantage of transferring a negligible recoil to the $^{229}$Th atom/ion, compared to the rather violent nuclear decay processes encountered in $^{233}$U decay. On the downside, interaction of any sample with a strong 29 keV X-ray beam produces a strong background and significant damage.

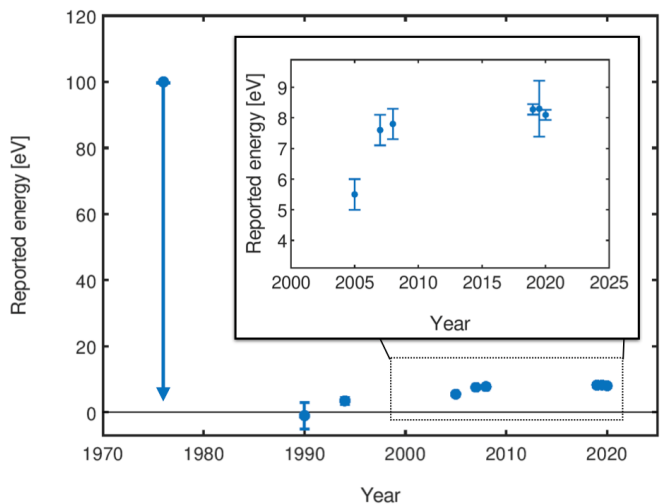

**Fig. 2.11:** Historical overview of the reported excitation energy of the $^{229m}$Th isomer.

To conclude this section on the present knowledge of the isomer properties, the remarkable progress achieved in recent years (see Fig. 2.11 for a historic overview of the reported excitation energy of the $^{229m}$Th isomer) now paves the way towards the realization of a nuclear clock. In particular, the much-improved determination of the isomeric excitation energy allows for assessing the appropriate laser technology for the long envisaged optical control of the $^{229m}$Th clock transition. With a VUV transition wavelength around 150 nm at least presently no CW laser is available to drive a nuclear clock, so other concepts will be pursued as discussed in the following section.

## 3 Towards the realization of a nuclear clock

Worldwide today vivid efforts are made to unravel the properties of the thorium isomer and to realize the nuclear clock: At UCLA, experimental efforts are focused on constructing a solid-state optical clock based on the $^{229}$Th nuclear transition via finding and constructing a suitable host crystal and determining the clock transition frequency [44,45,84]. At JILA/NIST activities concentrate on the development of a VUV frequency-comb laser able to drive the $^{229m}$Th clock transition, on concepts for direct frequency-comb spectroscopy of $^{229m}$Th targeting an internal-conversion-based solid-state nuclear clock [85,86]. At the SPring-8 synchrotron light facility in Japan X-ray pumping into the thorium isomer has been realized [79] and in Moscow methods are developed for the production and trapping of thorium ions for nuclear transition investigation [87,88].

The main objectives of ongoing and upcoming efforts towards the realization of a $^{229m}$Th based nuclear clock and its application, in particular as a quantum sensor for new and fundamental physics, comprise (i) to precisely determine the $^{229}$Th nuclear structure parameters to quantify the sensitivity of the nuclear clock to variation of fundamental constants of physics, (ii) to improve the precision on the nuclear transition frequency by about 9 orders of magnitude (< kHz level) to enable nuclear clock operation and feedback laser stabilization and (iii) to perform comparative measurements between different nuclear and atomic clocks to probe fundamental concepts of physics. To reach these ambitious goals, a distributed, 3-step research strategy which we describe in this review will be adapted amongst the teams forming the 'ThoriumNuclearClock' consortium within a 'Synergy' project framework of the European Research Council (ERC) [62].

A first series of experiments will start from the current knowledge of the $^{229}$Th nuclear and electronic structure parameters and advance this knowledge to a level where a) direct laser spectroscopy can be commenced, and b) the sensitivity of the nuclear transition energy to variations of fundamental constants of physics can be quantified (see more detailed discussion in Sec. 4). The second phase will implement direct resonant or near-resonant (electronic bridge) excitation of the $^{229}$Th isomer by laser light and determine the transition energy to GHz precision. The third phase will employ frequency comb precision spectroscopy to kHz level and laser stabilization (clock operation) on the transition, together with a systematic study of shifts and broadenings (clock benchmarking). This will continuously lead over into clock comparison experiments (with primary standards and an Yb$^+$ optical ion clock at the site of the German metrology institute PTB, via GNSS-based time transfer equipment and the portable Yb$^+$ Opticlock [89] at the sites of LMU Munich and TU Vienna) for fundamental physics tests and searches for variations of fundamental constants.

Experimental efforts will be continuously updated and improved by sustained theory-experimental collaboration. Theory work will a) guide the extraction of the nuclear properties and sensitivity of the nuclear transition energy to variations of fundamental constants b) provide critically needed atomic properties of Th$^+$, Th$^{2+}$, and Th$^{3+}$ ions for electronic bridge excitations c) develop ideas and measurement protocols for fundamental physics tests and dark matter searches with the nuclear clocks, including tests requiring 3 clocks at different sites (TU Vienna, LMU Munich, PTB Braunschweig).

## 3.1 Advancing knowledge of key nuclear properties of $^{229m}$Th

The experiments described in this subsection will deliver the nuclear transition energy sufficiently precise (target uncertainty <1 nm, corresponding to 0.05 eV or 10 THz) to start searches for laser excitation with GHz resolution. Isomer lifetimes for different charge states, nuclear-electron couplings and (ratios of) nuclear moments will be determined sufficiently precise to gauge the achievable sensitivity to fundamental constant variations and propose and plan experiments testing fundamental concepts of physics. Primarily the $\alpha$-decay of $^{233}$U and recoil ion sources will be used as "isomer factory" as described above, with a confirmed [60,61,71] branching ratio of 2% into the isomer state. Populating the isomer through $\beta$-decay of $^{229}$Ac (branching into isomeric state unknown, but estimated to >13% [18]) will be introduced as a new approach, transferring significantly less (~3 eV compared to 84 keV) energy to the $^{229}$Th nucleus. Controlled excitation to the isomeric state through off-resonant processes will be studied: pumping of the isomer through the 29 keV 2$^{nd}$ excited nuclear state as recently demonstrated by a collaboration between TU Vienna and Spring-8 in Japan [79]. Schemes of (laser-induced) electronic bridge processes in trapped ions will be investigated. All these experiments envisaged for the next round of experimental campaigns can be performed temporally in parallel to the VUV laser developments preparing for direct laser excitation of the thorium isomer (see Sec. 3.2).

### 3.1.1 Refined measurement of the transition energy

Complementary experimental approaches are presently being pursued and optimized to further refine the knowledge of the isomeric excitation energy before the first laser spectroscopic excitation measurements will be realized.

Within the gamma spectroscopy approach, rapid progress is currently being made in improving the detector resolution, thanks to the emergence of cryogenic detectors (microcalorimeters, transmission edge, nanowire detectors…). We expect that the resolution can be improved down to 3-5 eV within the next few years. This should allow to further reduce the uncertainty on the isomer energy to 0.05 eV. As discussed in Sec. 2.2, further progress can also be made by improving the absolute energy calibration of the spectrometers using new reference lines. While previous analyses focussed on individual datasets, a new field of comparative gamma spectroscopy is emerging, combining the existing (partially historic) data to gain new insights. Other approaches are discussed below, while plans for electronic bridge experiments are presented in Sec. 3.2.5.

#### 3.1.1.1 IC electron spectroscopy

In order to realize the isomeric excitation energy measurement via IC electron spectroscopy [66], neutralization of $^{229m}$Th ions was initiated while traversing a thin graphene foil. However, an alternative scenario is to impinge the ions onto a metallic surface and collimate, transport and register the resulting IC electrons behind a grid with a variable blocking potential. By applying retarding fields, an integrated spectrum can be generated, where all electrons with an energy sufficient to pass the retarding fields that are applied to high-transmission grids are counted.

In [90] the feasibility of this method regarding electron rates and spectral shape was studied in detail via simulations. The energy distribution of the electrons will reflect the electronic structure of the catcher's surface. Using an electron spectrometer in electrical contract with the catcher surface, the work functions of the catcher surface ($W_C$) and the one of the spectrometer ($W_S$) have to be considered, while their Fermi levels $E_F$ will align. A contact potential difference $\Delta E = W_C - W_S$ will be generated between the local vacuum levels. This situation is illustrated in Fig. 3.1a). If the work function $W_S$ of the spectrometer exceeds the catcher's work function $W_C$, an offset voltage $\Delta U$ needs to be applied to the catcher surface in order to provide the electrons with enough kinetic energy to overcome the contact potential difference. Consequently, the contact potential difference is shifted by $\Delta U$ (see Fig. 3.1b): $\Delta E' = W_C - W_S + \Delta Ue$.

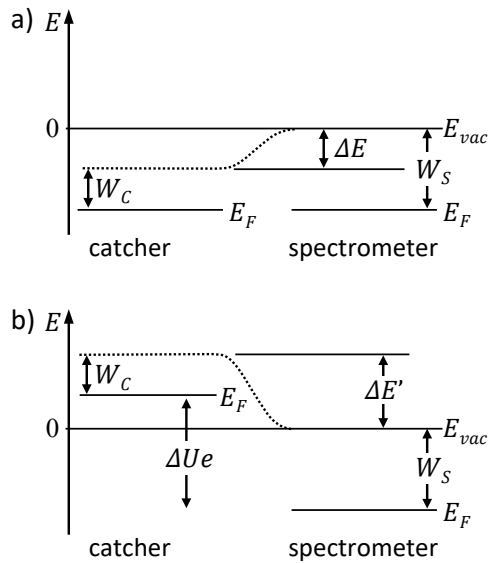

**Fig. 3.1:** Visualization of the work function of the catcher and spectrometer (see also [72]). Two situations are shown: a) Contact potential difference $\Delta E$ generated between the catcher and the spectrometer. b) Influence of an applied offset voltage $\Delta U$ on the contact potential difference. From [90].

Photons with an energy $h\nu \geq W_C$ can release electrons from a metallic surface with work function $W_C$, resulting in a Fermi distribution for the kinetic energies of these photo electrons with a maximum energy of $E_C^{max} = h\nu - W_C$. While this energy is given relative to the local vacuum energy level of the catcher surface, the maximum energy of the electrons registered with the spectrometer amounts to $E_S^{max} = h\nu - W_S + \Delta Ue$. One of the main results found in [90] was that the finally registered energy of the electrons does not depend on the value of the work function of the catcher surface, but only on the spectrometer work function and the applied offset voltage $\Delta U$. Treating the isomeric IC decay of $^{229m}$Th as a virtual photon with energy $E_I = h\nu$ that is coupling to the delocalized surface valence electrons in the catcher surface, the excitation energy of the isomer $E_I$ can be inferred from $E_I = E_S^{max} + (W_S - \Delta Ue)$, where the term $(W_S - \Delta Ue)$ can be measured with a light source of known energy and using the above expression for $E_S^{max}$. Therefore, the only remaining surface influence of the catcher surface on the maximum kinetic energy of an electron is the temperature-dependent Fermi distribution of $E_e^{max}$, but not the value of the surface work function. Nevertheless, the work function of the catcher surface still plays a role, since $h\nu = E_I \geq W_C$ must always be satisfied and the electron energies (with respect to the vacuum level of the sample) are distributed between 0 and $(E_I - W_C)$.

Improved measurements with the magnetic-bottle electron spectrometer that was described in Sec. 2 will address the integrated kinetic electron energy spectra of IC electrons emitted from different metallic surfaces with different work functions (Ta: 4.19 eV, Ni: 5.0 eV, Pt: 5.32-5.66 eV, Au: 4.8-5.4 eV). Previous measurements of this type were limited by the sensitivity around the end point of the kinetic energy spectrum (expressed by the number of registered electron counts as a function of the blocking voltage applied to the retardation grid of the electron spectrometer). In this spectral region lowest IC decay signal intensities compete with various noise contributions in the MCP signal down to a maximum sensitivity of 1 electron in 8 hours of measurement time. Sensitivity improvement will be realized by re-focusing the IC electrons behind the retardation grid onto a small-area MCP with considerably lower dark count rate compared to the previously used large MCP detectors. Also the now improved knowledge of the isomeric excitation energy [66,67] will allow to constrain the search range of the endpoint of the IC electrons' kinetic energy spectrum. Moreover, further improvement of the precision in determining the IC electrons' kinetic energies could be achieved by using a selection of metallic surfaces with a spread of work function energies. This will lead to different slopes of the integrated electron spectra, all expected to hit the energy axis at the same energy value of the maximum blocking voltage needed to repel the most energetic IC electrons. Improving the achievable precision of the $^{229m}$Th excitation energies by about a factor of 2-3 compared to the previous measurements is envisaged.

### 3.1.1.2 Direct radiative decay

Another promising route to determine the isomer energy more accurately is to detect the direct radiative decay that leads to emission of a "nuclear gamma ray" in the VUV range. The attainable energy resolution is determined by the spectral resolution of the used VUV spectrometers, usually in the range of 0.1-0.01 nm. Absolute calibration of the spectrometers is not critical, as a huge arsenal of reference lines is available with picometer accuracy.

These VUV spectroscopy schemes can be coarsely divided into two groups: Population of the isomeric state in nuclear decay processes and approaches which populate the isomer through excitation from the $^{229}$Th nuclear ground state, either resonant with the isomer or via pumping using higher-excited nuclear states (see Sec. 2.3 above).

Two nuclear decay processes are known to produce $^{229}$Th, with a certain probability in the isomeric state: The $^{233}$U $\alpha$-decay ($T_{\frac{1}{2}} = 1.6 \cdot 10^5$ years) has already been discussed extensively with a 2% feeding into the isomer. As an alternative, the $\beta$-decay of $^{229}$Ac ($T_{\frac{1}{2}} = 62.7$ min) can be used to produce $^{229}$Th, with a probability $14\% < \lambda_{\beta,\text{isomer}} < 94\%$ feeding the isomer [91]. Online production of $^{229}$Ac in the ISOLDE facility at CERN has been demonstrated and production via neutron activation of $^{228}$Ra→$^{229}$Ra (rapidly decaying to $^{229}$Ac in 4 minutes) is currently investigated [91].

Both processes are being used in schemes where $^{233}$U or $^{229}$Ac are doped or implanted into VUV transparent host materials [47]. Similar to the early attempts to detect the "VUV gamma ray" in spectroscopy [18–22], these experiments "simply" place a prepared sample in front of a VUV spectrometer. The route via $^{233}$U has the advantage of simplicity; doping of uranium into VUV-transparent materials (e.g. fluoride single crystals) is straightforward. This approach is however burdened with the violent α-decay producing the $^{229(m)}$Th ($Q_α$ = 4909 keV), transferring a recoil energy of 84 keV onto the daughter nucleus. This recoil energy is sufficient to overcome any chemical bond, ejecting the atom/ion from its initial configuration (e.g. dopant position in a solid) and leaving it in an uncontrolled final state concerning position and charge state. While this process inevitably generates a background signal, it may also open up additional decay paths, inhibiting the direct radiative decay of the isomer.

The approach employing $^{229}$Ac decay to populate the $^{229}$Th isomer has several advantages: it has a higher isomer production rate (at least 14%) and the β-decay ($Q_β$ = 1.1 MeV) transfers a small recoil energy of 2.3 eV onto the nucleus [92], below typical displacement energies in solids. Still, on-line production of $^{229}$Ac requires significant infrastructure and crystal damage due to the implantation will have to be considered.

The second main approach to "VUV nuclear gamma spectroscopy" relies on excitation paths starting from the $^{229}$Th nuclear ground state and leading into the isomer (possibly using intermediate states). Here, strong activities are under way to exploit the optical X-ray pumping demonstrated at SPring-8 (see Sec. 2.3 above), but now in a VUV transparent crystal sample. The main challenge here is to maintain the nucleus density of the solid-state sample, now in a transparent crystal sample. This required doping concentrations exceeding $10^{17}$ $^{229}$Th nuclei per cubic centimeter. Furthermore, the high-energy 29 keV X-ray beam causes significant damage and signal background.

Another obvious way to populate the $^{229}$Th isomer is by direct resonant radiative excitation. In view of the current lack of narrow-band laser sources in the VUV range, several attempts were made to trigger excitation using broadband synchrotron radiation. In 2014/15, a group at UCLA used $^{229}$Th-doped LiSrAlF$_6$ crystals ($10^{16}$-$10^{17}$ cm$^{-3}$ concentration) in the Advanced Light Source (ALS) synchrotron to optically excite the isomer [45]. ALS provided a 0.19 eV linewidth source with an integrated flux of ~7·$10^{14}$ photons/s, tunable between 7.3 eV and 8.8 eV. No nuclear fluorescence signal could be detected, which allowed to exclude (at 90% confidence level) short isomer lifetimes (between 1-2000 s) in the scanned energy interval. This assumes that non-radiative decay processes are negligible in the crystal.

A similar experiment was carried out at the Metrology Light Source (MLS), this time using a $^{229}$Th-doped CaF$_2$ crystal ($10^{16}$ cm$^{-3}$ concentration) [46]. The energy range between 7.5-10 eV was investigated, and again no nuclear fluorescence could be observed. Also a surface-deposited sample of $^{229}$Th nitrate on a CaF$_2$ carrier substrate has been investigated at MLS in the energy range 3.9-9.5 eV, again not producing a result [48]. The failure to detect nuclear fluorescence in these three experiments raises suspicion on the presence of non-radiative decay processes in the crystal/solid-state environment. Further investigation on the electronic structure of $^{229}$Th in different chemical environments are required [46].

### 3.1.2 Isomer lifetime

Refined measurements of the isomeric lifetime, presently under preparation, will target two aspects: aiming at (i) deeper insight into the influence of the electronic environment on the lifetime of neutral $^{229}$Th isomers and (ii) realizing a first measurement of the ionic $^{229m}$Th lifetime.

#### 3.1.2.1 Surface influence on the $^{229m}$Th lifetime

The so far only experimentally determined value of the thorium isomer's half-life is the value of 7(1) μs determined in [68] for neutral surface bound isomeric $^{229}$Th atoms. However, this value has to be treated with some caution, as it has been obtained after initiating IC decay of $^{229m}$Th ions via neutralization on a metallic surface, in this case an MCP detector with nickel alloy surface. Since the lifetime of the isomer depends on the electronic orbitals in the surrounding of the nucleus, one has to expect that the isomeric lifetime is affected by the chemical structure of the surface [31]. This was already observed in early lifetime measurements of the nuclear isomer in $^{235}$U [93,94] (which is the energetically second lowest nuclear excited state after $^{229m}$Th). There it was found that the lifetime of the internal-conversion decay channel on a catcher surface is influenced by the host material.

In [68] this was taken into account by the error margin of 1.0 μs. However, in subsequent measurements Pt was used as an alternative metallic catcher surface to initiate IC decay of $^{229m}$Th by neutralizing impinging ions and considerably modified half-lives were observed.

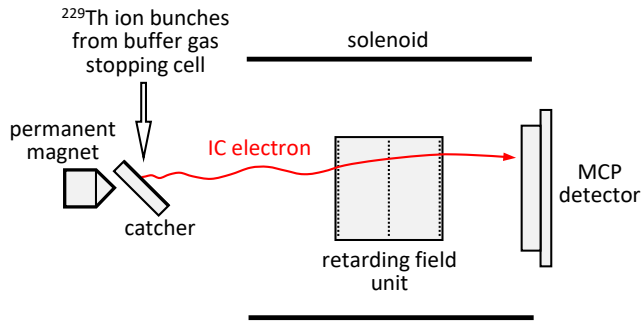

**Fig. 3.2:** Sketch of the experimental setup that was used to measure the decay properties of $^{229m}$Th in different host materials. $^{229m}$Th ion bunches are collected on the surface of a catcher which is placed in the collection region of a magnetic bottle spectrometer [69].

These measurements were performed with $^{229}$Th$^{2+}$ ions which were collected on a platinum surface with a kinetic energy of 60 eV: the 2+ ions started at a potential of +15 V in the RFQ-buncher and were collected on the catcher by an attractive potential of −15 V [69]. The corresponding experimental arrangement, based on the already introduced magnetic-bottle electron retardation spectrometer, is shown in Fig. 3.2. It was found that the measured lifetime depends on the applied blocking voltage. Additionally, it turned out that for larger blocking voltages the lifetime decreased. Modifications by up to a factor of 2 were observed [69]. These findings are interpreted such that the decay of the isomer was governed by more than one decay constant, resulting from different electronic environments. Besides surface effects introduced by water or hydrocarbon contamination layers on the untreated metal surface, also the penetration depth of the $^{229m}$Th ions into the platinum material could have been responsible for the observed behaviour of the isomeric half-life. The ions' initial kinetic energy of 60 eV results in an implantation depth of ~5 Å. IC electrons that are emitted deeper into the catcher material will lose kinetic energy by inelastic collisions with the catcher material. Such electrons can be suppressed more easily by increasing the blocking voltage of the magnetic bottle spectrometer. Hence electrons reaching the spectrometer with higher kinetic energies are more likely to originate from areas at or near the catcher surface, while the low-energy electrons are mostly due to scattered electrons originating from deeper-seated areas of the catcher.

Further systematic measurements with controlled clean surfaces of different metals and variable kinetic energies of the impinging $^{229m}$Th ions are needed to further investigate the dependencies of the isomeric lifetime.

### 3.1.2.2 Determination of the ionic $^{229m}$Th lifetime: instrumental prerequisites

So far, the measurement of the ionic lifetime of $^{229m}$Th was inhibited by the expected long lifetime of a few $10^3$ seconds, requiring correspondingly long storage times in the applied linear Paul trap. While a conventional high-vacuum environment did not allow for storage times of more than about 1 minute, the ionic isomeric lifetime needs storage times that can only be achieved in a cryogenic environment. Moreover, a measurement of the expected long lifetime will require an appropriately adapted methodology, different from the one used for the 9 orders of magnitude shorter lifetime of neutral $^{229m}$Th. Therefore, major instrumental developments are needed to prepare for the first determination of the half-life of ionic $^{229m}$Th.

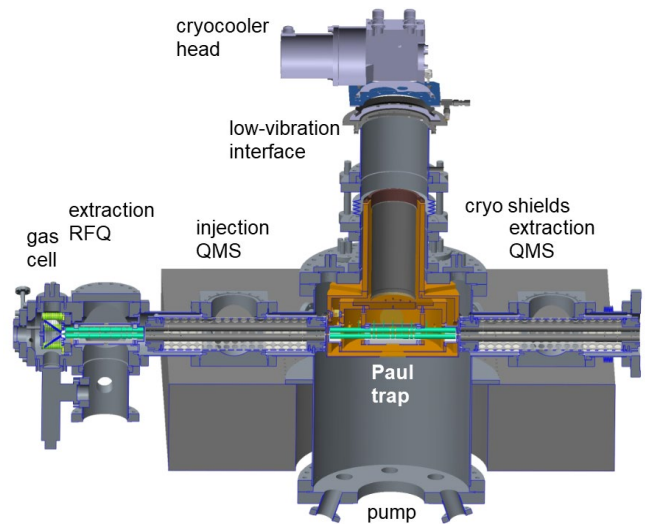

**Fig. 3.3:** Design sketch (side view) of the new experimental platform for laser-spectroscopic studies of stored thorium ions. A newly designed compact buffer-gas stopping cell containing the $^{233}$U source and an RF+DC funnel structure (left side) is followed by an extraction radiofrequency quadrupole (RFQ) and a quadrupole mass separator (QMS). The central part is formed by the cryogenic linear Paul trap, embedded into 4 K and 40 K cryoshields and cooled to 4 K by a vibrationally decoupled cryocooler head. A number of 12 viewports grant optical access to the trap center for diagnostics and laser manipulation purposes. A second QMS is mounted on the extraction side of the Paul trap, enabling a mass-selective diagnostic behind the Paul trap and the injection of Sr$^+$ ions for sympathetic laser cooling as described in the following.

The key element will be a cryogenic linear Paul trap, providing a platform for long ion storage, laser cooling and laser manipulations of stored thorium ions in the isomeric state. The system under development at LMU Munich builds on the experience gained with the CryPTEx cryogenic Paul trap developed by the Heidelberg group for studies of highly charged ions [95,96]. Fig. 3.3 shows a sectional view of the full setup, from the injection line with a newly developed compact buffer-gas stopping cell, cooler-buncher RFQ and injection quadrupole mass separator (left side of the Paul trap) and a second QMS on the extraction side of the Paul trap. A cryocooler, decoupled from the trap by a low-vibration interface, serves to reach the operational temperature of 4 K at the Paul trap electrodes, thermally shielded by 4 K and 40 K cryoshields.

Fig. 3.4 displays the trap electrodes with their support and electrical contact rods (left), mounted (after gold-plating) into the innermost cryoshield (right).

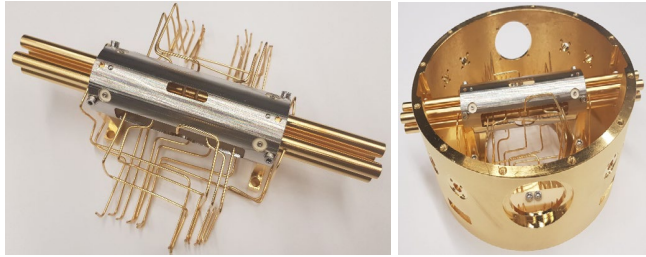

**Fig. 3.4:** Left: Gold-plated Paul-trap electrodes with surrounding support structure and electrical contact rods. Right: electrode system embedded into the 4 K cryoshield with diagnostic and laser entrance ports.

Fig. 3.5 presents the cryotrap setup with injection and extraction components (on the left-hand side the new gas cell is not yet mounted) as sketched in Fig. 3.3. The system has been commissioned with stable test ions, reaching a mass resolving power (tested with $^{39,41}$K and $^{85,87}$Rb) of $m/\Delta m \approx 150$ [97].

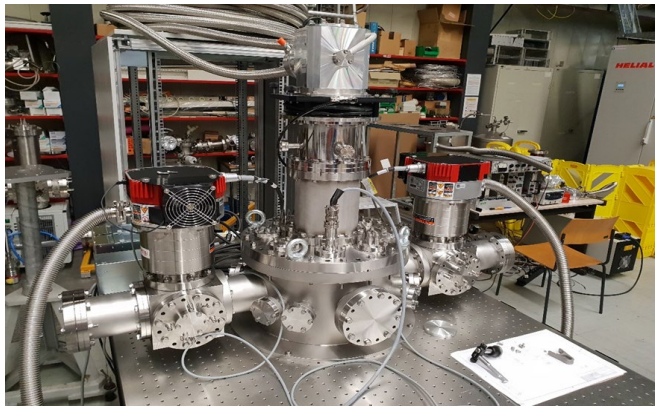

**Fig. 3.5:** Cryogenic Paul-trap setup with injection (left) and extraction (right) components as presently being commissioned at LMU Munich.

As only few $^{229m}$Th ions at a time will be loaded into the cryogenic Paul trap for sympathetic laser cooling and laser manipulation, an adapted compact recoil source and buffer-gas cell design was realized specifically for the new cryotrap setup. Compared to the stopping cell shown in Fig. 2.2, the new gas cell volume (displayed in Fig. 3.6) has been reduced by about a factor of 20 from a previously oversized volume (due to a different initial design application) of about 15000 cm$^3$ to now about 750 cm$^3$, sufficient to thermalize and extract the $\alpha$-decay recoil products from $^{233}$U with about 84 keV kinetic energy. Moreover, as only a few to a few hundred ions are needed to be loaded into the Paul trap, the $^{233}$U source can be reduced in activity from previously 290 kBq to about 10 kBq, thus requiring a much smaller diameter of only about 25 mm (compared to 90 mm in the cell from Fig. 3). Note that all $^{233}$U sources carry a central hole of 5 mm diameter in order to allow for collinear laser manipulations.

Cleanliness has been given highest priority in the design of the UHV compatible cell, allowing for bake-out up to about 150°C via heating elements inserted into the cell corner walls.

This system will form the central experimental platform for use in all envisaged trapped ion spectroscopy experiments.

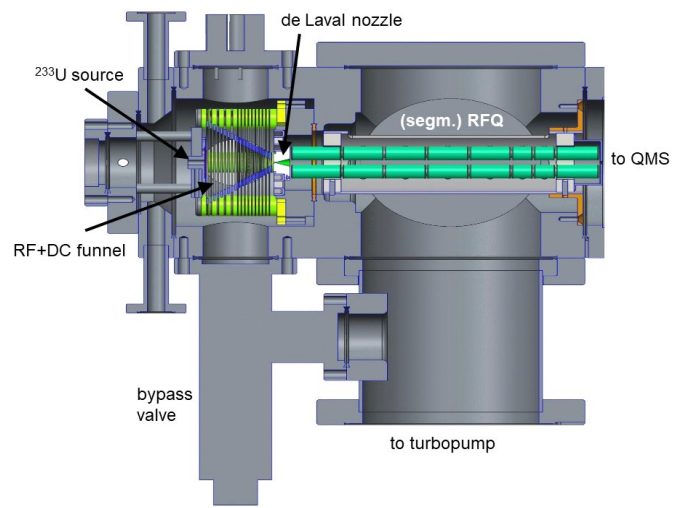

**Fig. 3.6:** Technical drawing of the newly developed compact $\alpha$ recoil source designed for loading the cryogenic Paul trap with $^{229(m)}$Th ions. The buffer-gas stopping cell (containing the $^{233}$U source and an RF+DC funnel structure) is connected to the segmented radiofrequency quadrupole (RFQ) by a de-Laval nozzle.

### 3.1.2.3 Measurement of the ionic charge- and electronic-state dependent isomeric lifetime

Following successful commissioning of the new experimental device comprising the compact gas cell and the cryogenic Paul trap, in a first stage towards an ionic lifetime measurement $^{229}$Th$^{3+}$ ions will be sympathetically laser cooled to mK temperature with co-trapped laser-cooled $^{88}$Sr$^+$ ions. These are injected via the extraction side into the Paul trap, through a quadrupole bender from an ion source mounted perpendicular to the trap axis to avoid a direct line of sight between the hot source and the cryogenic trap. $^{88}$Sr$^+$ is free from hyperfine structure and can be conveniently laser cooled with two diode lasers at wavelengths 422 nm and 1092 nm. This allows for preparing a large number of cold Th$^{3+}$ ions and to use them for spectroscopic and lifetime studies. Since the mass-to-charge ratio of $^{88}$Sr$^+$ is slightly higher than that of Th$^{3+}$, the two species remain closely coupled and big Coulomb crystals can be formed in a linear radiofrequency trap with Th$^{3+}$ ions close to the axis of the trap, inside a shell of cold Sr$^+$ ions. Fluorescence detection from each Th$^{3+}$ ion on the resonance transition at 690 nm with an intensified CCD camera will make it possible to distinguish between ground state and isomeric ions by illuminating the Coulomb crystal alternately with two laser frequencies that are tuned to an HFS component of one or the other state. This method makes it possible to detect small nuclear transition rates ($\ll$ 1/s) with high sensitivity in reasonable measurement times.

Moreover, the isomeric lifetime will be investigated in different charge states (Th$^+$, Th$^{2+}$, Th$^{3+}$) and selectively excited metastable electronic states, indicating contributions from electronic bridge decay. Hence detailed insight can be obtained into the ionic $^{229m}$Th lifetime as the last key nuclear property that has not been directly determined so far.

### 3.1.3 Determination of isomer and ground nuclear moments

While first results on the isomer nuclear moments have recently been obtained in two-photon laser spectroscopy of room-temperature Th$^{2+}$ [71] (see Sec. 2) with 15-30 MHz uncertainties in the hyperfine constants $A$ and $B$, significant improvements in accuracy can be expected from measurements with laser cooled Th$^{3+}$ ions. The electronic ground state manifold of Th$^{3+}$ consists of a doublet of 5F states with a total of 12 hyperfine levels in the nuclear ground state and 8 in the isomer. To obtain cyclic laser excitation and fluorescence detection in the presence of optical pumping requires a set of several laser frequencies that can be produced as RF modulation sidebands on a smaller number of laser carrier frequencies. The $^{229}$Th ions are kept at a mK temperature by sympathetic cooling inside a Coulomb crystal of $^{232}$Th or another coolant ion. Such a scheme has been successfully used in a hyperfine spectroscopy experiment for the nuclear ground state and $A$ and $B$ hyperfine constants with uncertainties of 0.6-10 MHz have been obtained [55] providing the basis for the so far most precise $^{229}$Th nuclear moments [98]. The experimental uncertainty has been due to uncertainties in Zeeman splittings combined with optical pumping effects, laser cooling dynamics, and light shifts [55]. An improved accuracy is expected for a planned experiment at PTB by employing methods that are used in microwave clocks based on hyperfine resonances in trapped ions, like driving the hyperfine transition with a microwave field that is temporally separated from the optical pumping in order to avoid light shift. The experiment will use a recoil ion source so that both the nuclear ground state and the isomer can be studied. Since these experiments may directly provide precise ratios of the nuclear moment of isomer and ground state without requiring information on the atomic structure, they are well suited to answer the question how similar or different the proton charge distribution is in both states (see Sec. 4).

It is interesting to note that nuclear moments, in particular the quadrupole moments of the ground and isomeric state, can also be studied in a solid-state environment. Doping of $^{229}$Th into VUV-transparent crystals – mostly fluorides – has been discussed quite extensively in the literature [99–101]. In most cases, the Th ion replaces the metallic ion in a different charge state, so charge-compensation mechanisms (vacancies, interstitial F ions) will take place. This breaks the crystal symmetry and leads to an electric field gradient of order 10-500 VÅ$^{-2}$ at the position of the nucleus, which in turn leads to a quadrupole splitting of the nuclear levels (see Fig. 3.7). Transitions between nuclear quadrupole states are expected in the 50-500 MHz range and can be investigated using nuclear quadrupole resonance spectroscopy in the radio frequency (RF) domain. Investigations in this direction are currently under way, starting with more abundant isotopes of similar quadrupole moment $Q$ such as uranium or neptunium. Nuclear quadrupole structure can on principle also be studied in Mössbauer spectroscopy, but the $^{233}$U→$^{229}$Th system is not well suited, due to the very weak gamma lines involved.

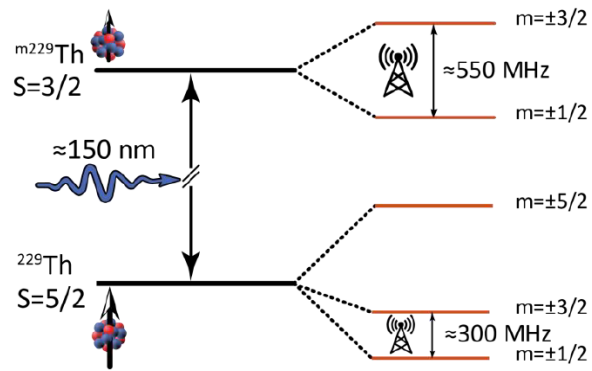

**Fig. 3.7:** Schematic of nuclear quadrupole splitting emerging for $^{229}$Th ions doped into VUV transparent CaF$_2$ crystals, with approximate transition frequencies in the RF regime.

### 3.1.4 Calculation of Th$^+$, Th$^{2+}$ and Th$^{3+}$ atomic properties

The determination of the ground/isomeric state nuclear moments requires a combination of precision measurements of the hyperfine structure and a precise calculation of the relevant hyperfine matrix elements. The planned experimental efforts have been summarized in the previous section. Present most accurate values of the $^{229}$Th ground state nuclear magnetic dipole and (spectroscopic) electric quadrupole moments, $\mu$ = 0.360(7) $\mu_N$ and $Q_s$ = 3.11(6) $eb$ [98], were obtained from the Th$^{3+}$ calculations using the linearized coupled-cluster method [102] combined with prior measurements [55]. The atomic hyperfine constants $A$ are proportional to the nuclear magnetic dipole moment $\mu$. Therefore, calculation of $(A/\mu)_{th}$ allows to extract $\mu$ as the ratio of the experimental value for the hyperfine constants $A_{expt}$ and $(A/\mu)_{th}$. The nuclear (spectroscopic) electric quadrupole moment $Q_s$ is obtained as the ratio of the experimental hyperfine constants $B_{expt}$ and the computed quantity $(B/Q_s)_{th}$. Since the accuracy of the resulting nuclear moments is limited by the estimated accuracy of the theoretical calculations, one has to select levels where the highest theoretical accuracy may be achieved. We note that nuclear moments can be extracted from computations and measurements for any Th ion. However, Fr-like $Th^{3+}$ ion has a single valence electron above the closed $[Rn]=[Xe]4f^{14}5d^{10}6s^26p^6$ core and such electronic level structure is generally conducive to highest accuracy atomic calculations. The uncertainty of the computations can be evaluated based on the differences in various approaches and comparison of hyperfine constants for similar elements with experiment, where magnetic moments are known to high accuracy from other methods [98]. The accuracy of the $Th^{3+}$ computations may be improved with inclusion of triple and non-linear excitations into the computations, such as performed in [103].

Th, $Th^+$, $Th^{2+}$ atomic properties were calculated in [104] using a hybrid method that combines the configuration interaction (CI) approach with the linearized version of the coupled-cluster method (referred to as CI+all-order method [105]). In this approach, a coupled-cluster method is used to build an effective Hamiltonian that includes core correlations. This effective, rather than bare Hamiltonian, is then used by the CI. Ref. [106] demonstrated that the CI+all-order method can accurately describe isotope shifts in such complicated systems as $Th^+$. The results of [106] are important for the selection of efficient transitions for the excitation of the $^{229}$Th isomer, as the isotope shift indicates the strength of the Coulomb interaction between electrons and nucleus that is relevant for NEET processes.

Isotope shift calculations in $Th^+$ and $Th^{2+}$, which included the specific mass shift, were carried out in [73] using the CI-order method. Theoretical results were combined with experimental measurements to extract the difference in *rms* radii between $^{229}$Th and $^{232}$Th. Recent advances in the development of the CI+all-order approach allow to significantly increase the number of configurations that can be included in the CI with new parallel code [107] and extract many more levels that was previously possible [108]. Triple and non-linear terms can also be added to the construction of effective Hamiltonian.

In summary, a wide variety of atomic properties of all Th ions and neutral Th can be computed to high precision to support experiments described in this work.

## 3.2 Nuclear laser excitation and spectroscopy

As much as recent experimental progress in our understanding of the thorium isomer's properties has opened the door towards the realization of the nuclear clock, the illustration in Fig. 3.8 reveals the enormous gap on the frequency scale between our present knowledge and the ultimate goal to operate a high-accuracy nuclear clock. The uncertainty of the $^{229m}$Th nuclear clock transition frequency is shown for different milestones on the road towards the nuclear clock. Starting from the present uncertainty of about 40 THz, still 12-13 orders of magnitude need to be bridged before arriving in the regime of metrological competitive clock operation.

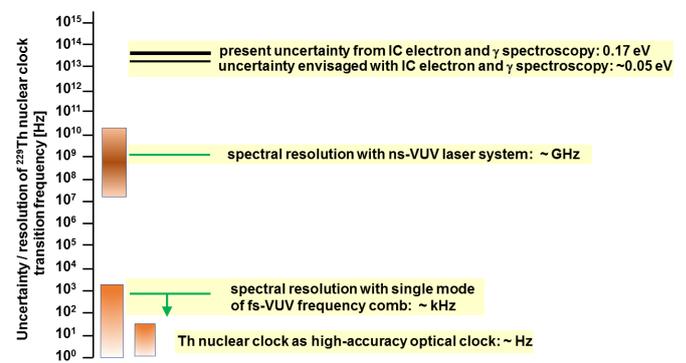

**Fig. 3.8:** Range of frequencies for the uncertainty (or resolution) of the $^{229m}$Th nuclear clock transition, from our present nuclear-physics based knowledge down to the ultimate goal of a high-accuracy optical nuclear clock operated by resonant laser excitation of the thorium isomer.

Within the 'ThoriumNuclearClock' project [62] it is envisaged to reach this final goal in three stages, starting with consolidating measurements that will improve the precision of the isomeric excitation energy via optimized $\gamma$ and IC electron spectroscopy to well below 0.1 eV. Searching for the energetic position of the thorium isomer's nuclear ground-state transition with laser spectroscopic precision marks the cornerstone of the second experimental phase. As proposed in [109], already existing laser technology can be applied for a proof-of-principle first all-optical resonant nuclear excitation of $^{229m}$Th. As will be outlined in the following section, using a tunable laser system with a bandwidth of about 1-10 GHz will allow for a broadband search of the nuclear resonance.

### 3.2.1 Direct VUV laser excitation of trapped ions

Tunable sources of coherent radiation in the VUV can be obtained by sum- and difference-frequency conversion of pulsed laser radiation in gases. A light source using resonance-enhanced four-wave difference mixing in xenon gas delivers high-intensity pulses of VUV light with a continuously tunable wavelength from 122 nm (10.2 eV) to 168 nm (7.4 eV) [110], covering a $3\sigma$ search range around the present best values for the isomer energy. Here one laser photon at 250 nm is close to a two-photon resonance in Xe at 9.9 eV and the second photon from a tunable laser is used for sum- or difference frequency generation. Precise control of the emission wavelength can be obtained by using light from frequency stabilized CW lasers for seeding pulsed amplifiers. The VUV radiation intensity of the source is in the range of $10^{13}$ photons per 5 ns pulse for the wavelengths 150-160 nm. The linewidth of the VUV radiation is in the range of a few 100 MHz and the pulse repetition rate, determined by the pulsed amplifiers, in the range of tens of Hz. Typical photon spectral densities of VUV radiation at 150 nm obtained in the four-wave mixing experiments are ~$10^6$ photons/s/Hz which corresponds to ~1 pW/s/Hz spectral power density. Because of the high spectral brightness, these VUV light sources are a useful tool for VUV spectroscopy of thorium. The ~GHz linewidth is adapted to the Doppler-broadened linewidth of trapped ions at room temperature, and allows for the resolution of hyperfine structure splittings and isomer shifts in doped crystals. Such a source is presently under development at PTB and a simplified schematic is shown in Fig. 3.9. Assuming the spectral power density of the VUV source is in the order of $10^6$-$10^7$ photons/s/Hz and the beam waist of 100 µm, the excitation rate of the thorium isomeric state can be estimated to be about $10^{-6}$-$10^{-5}$ s$^{-1}$ for an ion in an RF trap. It will therefore be necessary to work with many ions in parallel and to use an efficient method for detection of the isomer excitation, like a double-resonance scheme based on hyperfine spectroscopy [71].

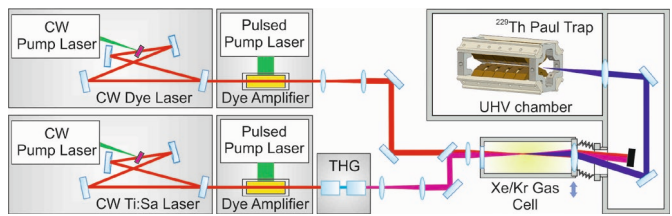

**Fig. 3.9:** Tunable VUV laser system based on four-wave mixing of amplified laser pulses in Xe or Xe/Kr gas. The pulsed amplifiers are seeded by tunable CW lasers for precise frequency control. The third harmonic of one laser (250 nm) is close to a two-photon resonance in Xe at 9.9 eV and the photon energy from the second laser is subtracted to reach the 8 eV region.

### 3.2.2 Broad-band resonant excitation of $^{229}$Th

Our present knowledge of the thorium isomer's excitation energy still requires to scan a considerable energy interval (at least 0.34 eV corresponding to the $1\sigma$ uncertainty of the isomeric energy from [66,67]) while searching for the precise energy of the nuclear resonance. In order to allow for a reduced laser scanning time and in contrast to previous laser experiments, this scenario makes use of the fast (lifetime of ~10 µs) non-radiative internal-conversion decay channel of neutral $^{229m}$Th for the isomer detection. A large signal-to-noise ratio can thus be achieved as prerequisite for a reasonably short laser scanning time. The IC decay channel of $^{229m}$Th has been experimentally firmly established [27,28] and is known to be the dominant decay channel for neutral $^{229}$Th atoms. This approach corresponds to laser-based conversion electron Mössbauer spectroscopy in the optical region [29] and is found to be advantageous compared to earlier proposals that make exclusive use of a potential radiative decay for isomer detection. The reason for this improved perspective is that the isomeric IC decay is ~9 orders of magnitude faster, which allows us to trigger the decay detection correlated with the laser pulses. In this way the signal-to-background ratio of the detection can be significantly improved, while shortening the required time to search for the direct nuclear laser excitation of $^{229m}$Th (across an energy range of 1 eV) to a realistically feasible period of about three days [109]. Fig. 3.10 displays a scheme of the experimental setup as developed in [109], where a thin layer of $^{229}$Th (2-3 nm thick, area ca. 1 mm$^2$) coated onto a gold substrate is irradiated by a broad-band (ca. 10 GHz), pulsed (pulse duration ca. 5 ns, repetition rate ca. 10 Hz) and tunable VUV laser (similar to the one described in [110]). IC decay is initiated upon resonant excitation of $^{229m}$Th, the emitted IC electrons are collimated by a permanent magnet and guided in a solenoidal magnetic field towards an MCP detector.

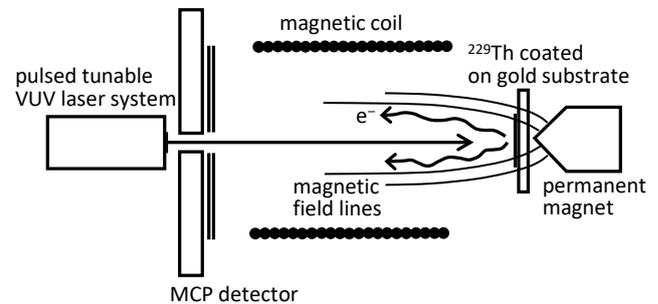

**Fig. 3.10:** Schematical view of the experimental scheme for direct laser excitation of $^{229m}$Th making use of the IC decay channel in the neutral thorium isomer [109].

Within the 'ThoriumNuclearClock' project, direct VUV laser nuclear excitation studies are envisaged to be performed in collaboration with PTB (Braunschweig, Germany) using their tunable four-wave mixing VUV laser system with coincident IC electron detection from an irradiation of solid targets according to the scheme proposed in [109].

### 3.2.3 Narrow-linewidth high-power VUV frequency comb

An advanced concept for a narrow-linewidth direct frequency-comb spectroscopy of $^{229m}$Th and an internal-conversion based solid-state nuclear clock was recently worked out in [86]. The concept uses a single comb mode of a vacuum-ultraviolet frequency comb generated from the 7th harmonic of an Yb-doped fiber laser system. More than $10^{13}$ $^{229}$Th atoms on a surface are irradiated in parallel and a successful nuclear excitation is probed via the internal-conversion decay channel. A net scanning time of 15 minutes for the 1σ energy uncertainty interval of 0.34 eV of [66,67] appears to be achievable when searching for the nuclear transition. In case of successful observation, the isomer's energy value would be constrained to an uncertainty of about 100 MHz, which is a factor of $10^6$ of improvement compared to today's knowledge. Further, the comb mode could be stabilized to the nuclear transition using the same detection method, allowing for the development of an IC-based solid-state nuclear clock, which is estimated to achieve the same performance as a crystal-lattice nuclear clock, however, with the advantage of a drastically simpler detection scheme [86]. Finally, in the concept of [86] the same laser system with a comb mode linewidth of 0.5 kHz could be used to narrow down the isomer's transition energy by further six orders of magnitude during laser excitation of $^{229}$Th$^{3+}$ ions in a Paul trap, and assuming a smaller comb mode linewidth of 1 Hz to drive nuclear Rabi oscillations, as required for the development of a nuclear clock based on a single $^{229}$Th$^{3+}$ ion.

A VUV frequency comb for the optical excitation of the nuclear transition in trapped $^{229}$Th ions needs to fulfil the requirements of the coverage of the search range in the VUV around 150 nm, a large power per comb mode (and average power), and a narrow linewidth of the comb modes.

The recent improvement in the determination of the $^{229m}$Th transition energy can be considered sufficient to choose a system with an ytterbium-based laser, reaching the VUV with the 7th harmonic. This allows for a higher IR average power compared to other systems, e.g. Ti:sapphire, reaching the VUV with the 5th harmonic. Yb:YAG and Yb:fiber amplifiers deliver multi-100 W average power, which overcompensates the expected larger conversion efficiency for shorter driving wavelengths [111]. An Yb:YAG Innoslab amplifier would provide a large power with very small nonlinearity in the amplifier at multi-10 MHz repetition rate even without chirped-pulse amplification (CPA) [112], however, the center wavelength is fixed at 1030 nm. Yb:fiber amplifiers allow a larger flexibility in the center wavelength and can reach a reasonable nonlinearity employing CPA. We plan to use a 400 W Yb:fiber amplifier operated with a center wavelength of 1050 nm, yielding a VUV spectrum centered at 150 nm. This power is 5× larger than the highest power for an IR frequency comb reported to date [113], and will allow scaling VUV power.

Another issue in scaling the VUV power is to increase the conversion efficiency by using shorter driving pulses. At the same time this increases the bandwidth in the UV, covering a larger search range. HHG efficiency is a steep function of pulse duration, which promises a larger VUV power per comb mode, even though the power is distributed over more comb modes. The driving pulse duration is limited by the bandwidth of the enhancement resonator, allowing pulses well below 100 fs depending on the finesse and number of mirror reflections. Even 19 fs have been demonstrated [114], however at moderate power and without HHG.

While Yb:fiber amplifiers achieve 120 fs pulses at the 80 W power level [113], the pulses are considerably longer at larger average power and nonlinear pulse compression has to be employed. Established compression schemes are not applicable to pulses with pulse energy of several μJ, however, which prevented the combination of the high average power of Yb-based lasers (>100 W) with short pulse duration (<100 fs) at high repetition rates (10-100 MHz). A novel compression scheme based on spectral broadening in a multi-pass cell fills the gap in applicable pulse energy [115,116]. Such a pulse compression setup has enabled a resonator-enhanced HHG system with a record power in the XUV [117]. The nonlinear pulse compression also adds some flexibility to the center wavelength. The nonlinear spectral broadening is symmetric around the center wavelength (with negligible Raman delay in a solid nonlinear medium), but the spectrum can be broadened by a large factor and part of the spectrum can be removed with a dielectric filter, shifting the center wavelength by >10 nm with an acceptable power loss.

An enhancement resonator will be used in order to reach the intensity for the highly nonlinear HHG process also a high repetition rate and to boost the conversion efficiency. Circulating average powers >10 kW in an HHG resonator and stable operation over many hours have been demonstrated [85,118]. At 150 nm a large output-coupling efficiency can be achieved with a coated plate at Brewster's angle [119]. We plan to use geometrical output coupling in a non-collinear configuration similar to [120], which avoids the dispersion, nonlinearity and degradation problems of the plate, and also promises a large output-coupling efficiency. The highest VUV power per comb mode reported to date is ca. 0.3 nW/mode at 149 nm [119], estimated from the out-coupled average power of 0.5 mW in the 7th harmonic. The scaled parameters of our planned system let us expect a VUV power well above 10 nW/mode.

A major concern for the development is the comb linewidth in the VUV, which is determined by the phase noise acquired in the system. If the phase noise becomes too large, the power in the comb modes drops and the and frequency comb structure eventually completely vanishes. The HHG process has been shown to be coherent, but the phase noise is multiplied according to the harmonic order, yielding a quadratic increase of the harmonic's linewidth [121]. This affords a sufficiently small phase noise of the IR frequency comb and in turn a careful design of the system. This includes a low-noise IR frequency comb stabilized to a high-finesse reference cavity in the optical domain, low-noise amplifiers and a compensation of beam path fluctuations. Also, the relative intensity noise (RIN) has to be considered, as it adds to the phase noise proportional to the nonlinear phase acquired in the fiber amplifier and nonlinear pulse compression. The enhancement resonator represents an essential component, as it suppresses high-frequency phase-noise contributions. Frequency combs in the VUV have so far only been applied to spectroscopy of broad transitions (>MHz) in argon and neon [122] and in xenon [123], not confirming narrow VUV comb linewidth but only placing an upper bound in the MHz range.

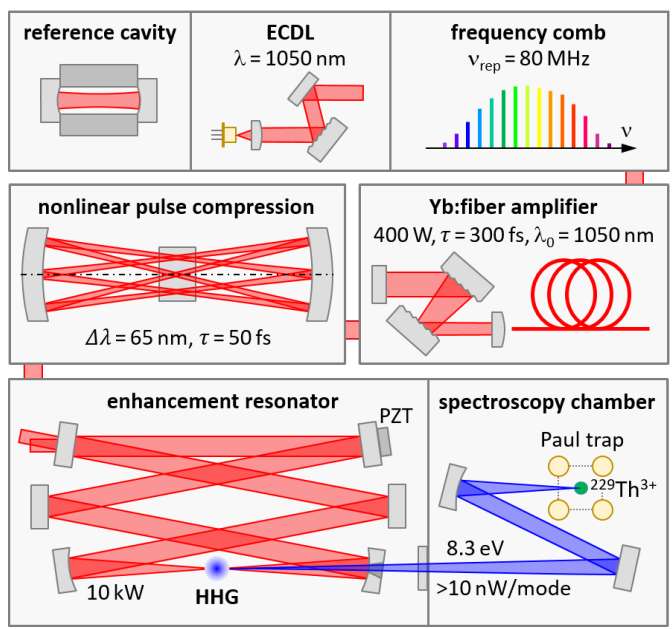

**Fig. 3.11:** Sketch of the planned VUV frequency comb with an IR frequency comb stabilized to a reference cavity via an external-cavity diode laser (ECDL), a high-power Yb:fiber amplifier, nonlinear pulse compression in a multi-pass scheme, and resonator-enhanced high-harmonic generation (HHG) in a gas target.

Such a laser system will be installed at LMU Munich. In combination with the previously described cryogenic Paul trap it will allow for direct resonant VUV laser excitation of (sympathetically laser cooled) trapped $^{229m}$Th$^{3+}$ ions.

### 3.2.4 Narrow-band solid-state VUV frequency comb

Very recently, generating high harmonics from solid-state material has attracted some attention [124–126]. The underlying physical processes are not yet fully understood and there is even a discussion, whether the dominating origin of solid-HHG is to be found within the bulk material or rather at the surface or at interfaces [126–128]. Still, this process promises high VUV conversion efficiencies and has significant practical advantages compared to gas-jet systems.

The team at TU Vienna recently demonstrated solid-HHG in a passive enhancement cavity, realizing a VUV frequency comb (see Fig. 3.12) [129]. The system is based on a commercial 0.9 W Ti:sapphire femtosecond oscillator (FC8004, Menlo Systems) generating 27 fs pulses at a repetition rate of 108 MHz, at 800 nm central wavelength. The repetition rate and the offset frequency of the frequency comb are stabilized to a 10-MHz rubidium frequency standard, realizing a frequency comb in the infrared (IR).

After a pre-compensation (negative chirp), infrared pulses with 7.5 nJ energy enter a secondary, passive enhancement cavity, built inside a UHV vacuum chamber to avoid absorption of the generated UV light. The group delay dispersions (GDD) of the cavity mirrors are chosen to yield effective near-zero GDD per roundtrip. The length of the ring-type enhancement cavity is 2776 mm to match the seeding laser repetition rate. One cavity mirror is mounted on a piezo to lock the cavity length to the repetition rate of the Ti:sapphire oscillator, using a Hänsch-Couillaud scheme. Without the solid-state target, enhancement factors ~200 are realized.

Inside the cavity, curved mirrors are used to focus the beam onto an IR-transparent solid-state target, with a FWHM focus of 12 μm. The target consists of a 100 μm sapphire carrier substrate with a 30 nm AlN crystalline film. The sample is tilted near Brewster angle to reduce losses inside the enhancement cavity (<1%, see Fig 3.12 (b)).

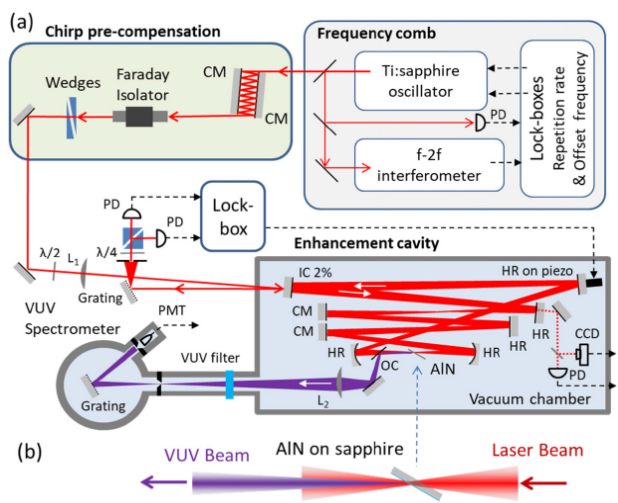

**Fig. 3.12:** Schematic of the solid-state VUV frequency comb system (from [129]).

The 5$^{th}$ harmonic (central wavelength at 160 nm) generated by the solid-state target are coupled out of the enhancement cavity using a multilayer mirror designed for >90% reflectivity within the 150-170 nm spectral range [130]. The 5$^{th}$ harmonic, together with weak residuals (few %) of the 3$^{rd}$ and 7$^{th}$ harmonic is focused onto the entrance slit of a VUV spectrometer using a lens. An additional VUV filter us used to suppress the otherwise overwhelming signal from the 800 nm driving radiation.

The 5th harmonic is generated efficiently from the solid-state target (see Fig. 3.13), also the 3rd and the 7th harmonic have been detected. Note that the relative amplitudes of the harmonics are strongly affected by the highly selective efficiency of the outcoupler. The frequency-comb structure of the generated UV light was verified by beating the 3rd harmonic with a 266 nm CW laser (Toptica TopWave 266). The beating signal also provided an upper bound on the linewidth of a single comb mode <1 MHz, most probably dominated by the linewidth of the CW laser.

We estimate the IR-VUV conversion efficiency of this process to be ~$3 \cdot 10^{-8}$ (for the 5th harmonic, referring to driving laser power outside enhancement cavity), similar to gas-jet efficiencies obtained in comparable systems [131]. Increasing the beam power circulating within the enhancement cavity by two orders of magnitudes (compatible with damage thresholds of involved optical components) and shifting the central pump wavelength towards 750 nm, we believe that a 5th harmonic centered at 150 nm with 0.1-1 nW/comb tooth can be realized. One option investigated at the moment in Vienna is to include the solid-HHG directly in the primary femtosecond oscillator. As the harmonics are derived from a Ti:sapphire primary frequency comb with rather low phase-noise and taking into account the quadratic scaling of the noise with harmonic orders, we believe, that a linewidth of few kHz for a single comb tooth can be achieved. This approach still produces 2-3 orders of magnitude lower spectroscopic yield than the high-power gas-jet approach described above (Sec. 3.2.3). It is nevertheless a promising approach for the interrogation of $^{229}$Th-doped crystal samples, where the number of interrogated nuclei is increased by 10 orders of magnitude (compared to trapped ion approaches) and where the linewidth of the interrogation (single comb tooth) of order kHz matches the (broadened) linewidth of the nuclear transition within the solid (equally kHz).

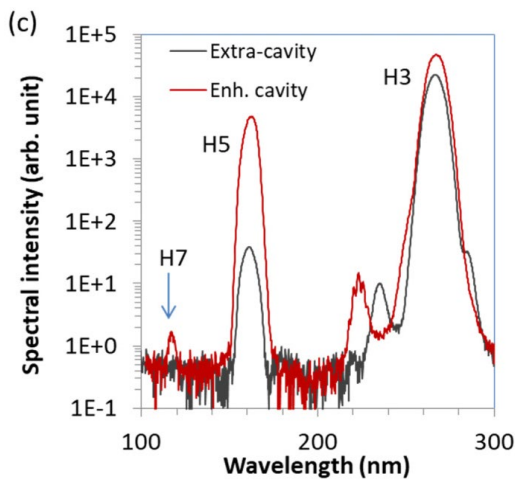

**Fig. 3.13:** High harmonics generated within the solid-state HHG approach. Black line indicates the single-pass yield, red line the yield obtained within the passive enhancement cavity (from [129]).

### 3.2.5 Electronic bridge excitation

We now shift from direct nuclear laser excitation scenarios towards a different setup involving the atomic shell as mediator. The electronic bridge (EB) is a process coupling a nuclear transition to a transition in the atomic shell via electromagnetic interactions. This becomes relevant especially in the case of low-energy transitions where the radiation wavelength exceeds the nuclear size by many orders of magnitude and radiative coupling to the larger electron shell serves as a somewhat better matched antenna. Luckily, the EB does not require a perfect energetic match between the atomic and nuclear transitions; the energy mismatch is covered by the emission or absorption of a photon. This process was first investigated theoretically in [132,133] and discussed in the context of $^{229}$Th in [134–137]. In the literature the term EB is used both for nuclear excitation and for nuclear decay, with the corresponding atomic shell transitions.

For illustration purposes we present a simplified three-step EB excitation scheme in Fig. 3.14. The atomic shell is initially in an excited state. Upon absorption of an external laser photon, a virtual electronic state is reached, which matches in energy the nuclear transition energy. In the final step, the electronic shell decays to a real state and transfers its energy to the nucleus, which is thereby excited.

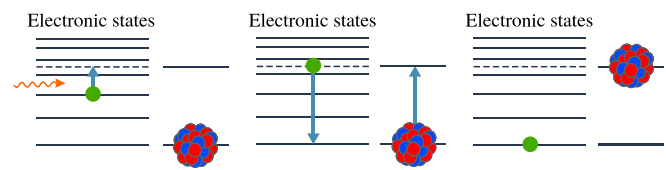

**Fig. 3.14:** A schematic illustration of the EB excitation scheme. The virtual state with an energy matching the nuclear excited state is depicted by the dashed line [138].

The theoretical formalism for the study of EB has been developed in Refs. [134–137] and relies on the calculation in third-order perturbation theory of the transition amplitude of the process. This amplitude involves the electronic and nuclear matrix elements of the respective transitions, and a sum over all possible intermediate states used for the description of the virtual electronic state. Should the virtual state be in the close vicinity of a real state, the corresponding term in the sum over intermediate states becomes very large and the EB process can be very efficient. Theoretical studies show that depending on the exact nuclear transition energy and on the ionic charge, the EB channel can be more efficient than photoexcitation or radiative decay, respectively [136,137]. This makes it appealing for the direct access to the isomeric transition in $^{229}$Th. It also offers the technical advantage that two-photon excitation can be employed by using an electronic intermediate state in a first excitation step, avoiding the need for a tunable VUV laser. EB excitation would likely by followed by EB decay, leading to a characteristic, delayed cascade of fluorescence photons. This signature has been searched for experimentally in Th$^+$ and Th$^{2+}$ ions, where the electronic spectra possess a high level density in the energy range of the isomer, making a resonant enhancement of the electronic bridge probable. For Th$^+$, more than 200 so far unknown electronic states of even parity have been observed between 7.3 and 9.8 eV [139], but an investigation of laser excitation of these levels has not provided an indication for nuclear excitation.

In the following we present three proposals to use the EB process for an enhanced precision in the energy determination of the $^{229m}$Th isomer and also as a means of efficient nuclear excitation. First, we review the prospects to use a laser-induced electronic bridge (LIEB) scheme for a stimulated decay of the nuclear isomer, with the possibility to precisely determine the nuclear transition energy [140]. While LIEB is not a nuclear excitation but a decay mechanism, it can lead to a refinement of our knowledge of the isomer energy as an additional approach to the ones discussed in Sec. 3.1.1. Second, we present an EB excitation scheme making use of highly charged ions which have a more advantageous level scheme as compared to the already investigated EB schemes in Th$^+$, Th$^{2+}$ and Th$^{3+}$ ions [141]. Finally, we discuss the possibility to use the EB mechanism in Th-doped VUV-transparent crystals such as CaF$_2$ [142].

### 3.2.5.1 LIEB

Nuclear de-excitation by LIEB can have a much faster rate than the radiative decay of the nucleus. LIEB is a version of EB process in which the additional photon is not emitted, but absorbed by the electronic shell from an externally applied laser field. The process is thus equivalent to the excitation of the electronic shell with two photons, one of which is provided by the laser source and the other one from the nuclear de-excitation [140]. A schematic illustration of the process is presented in Fig. 3.15. For the case of Th$^{3+}$ it has been shown that the nuclear decay in the LIEB process is four orders of magnitude faster than the radiative decay for the presently accepted value (combined from the two presently most precise measurements) of the isomer energy of 8.19(12) eV [66,67]. This in turn could be used for a more precise determination of the isomer energy.

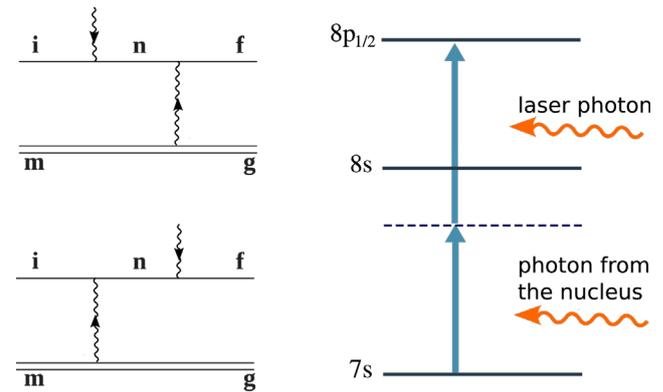

**Fig. 3.15:** Example of the LIEB process (right) and its Feynman diagrams (left). The solid lines in the Feynman diagrams correspond to the electronic states, the double-lines represent the nuclear state and the wiggly lines depict the photons [140].

The LIEB scheme requires production and trapping of $^{229m}$Th in the 3+ ionization state and the excitation of the electronic shell to the 7s initial state. Population of the isomer could be reached for instance by the 2% population channel in the $\alpha$-decay of $^{233}$U. The electronic shell can be efficiently transferred from the ground state to the 7s state making use of the stimulated Raman adiabatic passage (STIRAP) method [143]. Once reached, the 7s state has a long lifetime of 0.5 s allowing for the application of the LIEB scheme. Upon interaction with the tunable optical or ultra-violet laser, the isomeric state which otherwise has a long radiative lifetime of approx. $10^4$ s may decay via the LIEB process. Population and radiative decay of the upper electronic orbitals 8p or 9p could be observed by detection of fluorescence photons. The decay of these states occurs with emission of photons with energies in the range 15-20 eV, which could be easily differentiated from the laser photons used for excitation. The nuclear isomer energy $E_{is}$ can be found via scanning with a tunable laser for a LIEB resonance, i.e., for the population and decay of the upper state, and finally determined with a precision typical for laser spectroscopy.

### 3.2.5.2 EB in highly charged ions

EB schemes for excitation and decay have been investigated for Th$^+$ [137], Th$^{2+}$ [6], Th$^{3+}$ [136,140], and recently also for $^{235}$U$^{7+}$ [144]. A different idea to use highly charged ions for the EB excitation of $^{229}$Th has been proposed in [141]. Highly charged ions have a number of advantages compared to the low charge states previously investigated. On the one hand, when interacting with a laser, higher intensities can be used, since multiphoton ionization is not relevant. On the other hand, ion configurations with open d and f shells have a large variety of M1 transitions to match the isomeric transition. Based on these promising advantages, [141] has investigated EB schemes in Th$^{35+}$ ions. The particular level scheme is presented in Fig. 3.16.

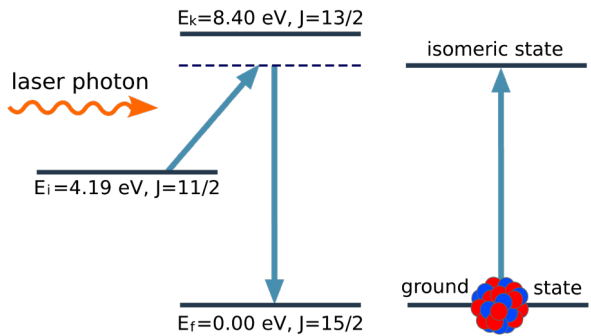

**Fig 3.16:** Simplified scheme of the EB excitation mechanism in Th$^{35+}$. The electronic shell (left) initially in an excited state $E_i$ is promoted by a laser photon to a virtual state (dashed line), which decays to a lower-lying real state $E_f$ by transferring its energy to the nucleus. The latter undergoes the transition from the ground to the isomeric state (right). Electronic states are labelled by energy and angular momentum $J$ [141].

The Th$^{35+}$ ion has an electronic configuration [Kr]4d$^{10}$4f$^9$. Provided that the isomer energy has a value within the determined 1$\sigma$ uncertainty range $E_I = 8.19(12)$ eV [63,64], a standard tunable UV laser could be used to excite the nuclear isomer via EB as shown in Fig. 3.16. The required excited initial electronic state can be quite efficiently populated in steady state by using highly charged ions produced in an electron beam ion trap (EBIT) [145]. The EB excitation would at the same time allow for the determination of the energy $E_I$ with a precision of $10^{-5}$ eV. Eventually, a successful EB scheme in Th$^{35+}$ should provide more efficient isomer excitation than direct VUV laser photoexcitation. Combined with suitable narrow-line electronic transitions with very low sensitivity to external perturbations (as expected in highly charged ions), this type of bridge mechanism could then potentially be used for the operation of a future nuclear clock.

### 3.2.5.3 EB in thorium-doped crystals

Rather than employing trapped ions, a more exotic version of EB can be applied in VUV-transparent crystals which have been doped with $^{229}$Th. For instance, CaF$_2$ has a confirmed band gap of 11-12 eV [146]. When doped with Th nuclei, a set of electronic defect states appear in the crystal band gap at energies close to the nuclear transition energy. What at first sight is rather a nuisance, can be used for an EB excitation scheme of the nuclear isomer [142]. The defect states provide a starting point for nuclear excitation via EB mechanisms involving stimulated emission or absorption using an optical laser.

A schematic illustration of the envisaged EB mechanism in Th-doped crystals in presented in Fig. 3.17. The initially populated electronic defect states $d$ lie in the crystal band gap (between the ground state $o$ and the conduction band $c$). At present, the defect states are only approximately known from theoretical predictions of density functional theory (DFT) [147] and their exact energy might lie above or below the isomer energy. In this case, the EB process occurs either spontaneously or assisted by an optical laser in the stimulated or absorption schemes, proceeding via a virtual electronic state.

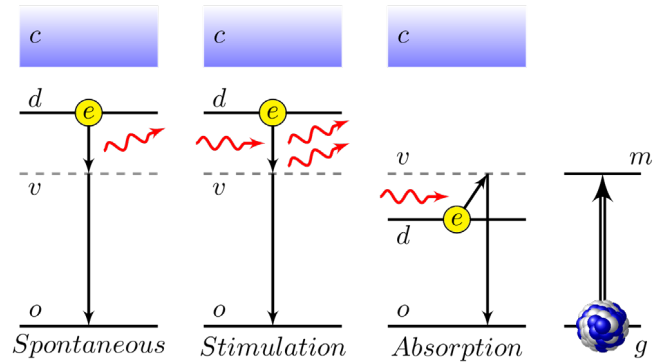

**Fig. 3.17:** EB process for the excitation of $^{229m}$Th from the ground state $g$ to the isomeric state $m$. The EB excitation can occur either spontaneously (left) or by stimulated absorption or emission (middle graphs) [142].

The rates of these processes calculated in [142] are at least 2 orders of magnitude larger than direct photoexcitation of the isomeric state using available light sources. This highlights the advantages of the crystal-EB approach. First, it exploits the electronic states of the crystal defects previously viewed as a nuisance, and second, it reduces the requirements on the VUV radiation to efficiently excite the isomer. In addition, the nuclear isomer population can also undergo quenching when triggered by the reverse mechanism, leading to a fast and controlled decay via the electronic shell. These findings are relevant for a possible solid-state nuclear clock based on the $^{229m}$Th isomeric transition.

### 3.3 Precision spectroscopy, clock operation, and fundamental tests

Investigations planned within the 'ThoriumNuclearClock' project will perform high-resolution, precision laser spectroscopy of the nuclear transition, converging to nuclear clock operation and first frequency comparisons between a nuclear clock and state-of-the-art atomic clocks, including primary Cs clocks for an absolute frequency measurement and trapped-ion optical clocks with Yb$^+$ as a test case for fundamental physics tests. Due to the potentially extremely high sensitivity of the nuclear clock to the variation of the fine-structure constant $\alpha$, at the level of $K \sim 10^5$, a clock-comparison experiment of the nuclear clock with any other atomic clock at the $10^{-15}$ uncertainty level, carried out in the span of one year will surpass several present limits by several orders of magnitude and have exceptional new physics discovery potential (see Sec. 4). Envisaged availability in this phase of the experimental efforts of three operational nuclear clocks at TU Vienna, PTB, and LMU Munich will result in unparalleled capability for testing fundamental new physics.

Using a single mode of harmonic radiation from a frequency stabilized femtosecond laser, a resolution around 1 kHz in the VUV is targeted. Since all systematic frequency shifts are expected to be much smaller, the obtainable uncertainty will be limited by statistics only. Two newly developed VUV frequency comb systems will be employed, to be located at LMU and constructed at TU Vienna (see Sec. 3.2.3 and Sec. 3.2.4). The latter is designed to be transportable for experiments at PTB. All frequency measurements at LMU Munich and TU Vienna will be referenced to local microwave oscillators that will be compared to the primary standards at PTB using time transfer via the global navigation satellite systems GPS and Galileo. Ultimately, long-term measurement series for clock comparisons (intra-nuclear and nuclear-optical) are foreseen. Also, a transportable single-ion optical clock based on Yb$^+$ developed at PTB in the framework of the Opticlock project will be operational, and may be employed for precision frequency comparison campaigns at LMU and TU Vienna.

### 3.3.1 High precision laser spectroscopy of $^{229}$Th$^{3+}$ and implementation of clock operation

The highest spectroscopic resolution and lowest systematic uncertainty are expected from laser-cooled $^{229}$Th ions in radiofrequency ion traps. The experimental methods have been refined and tested extensively on forbidden transitions in the electron shell [3] and are the basis of the presently most precise optical clock [4]. The clock ion of interest is laser cooled, either directly by driving one of its electronic resonance transitions or sympathetically by trapping it together with a coolant ion that provides such a transition if it does not exist in the clock ion. Both species are coupled by the Coulomb interaction and form a Coulomb crystal typically at a mK temperature. The fluorescence signal emitted on the cooling transition is strong enough to detect the presence, the state of motion and the internal state of a single ion. The narrow reference transition of the clock ion is then probed with radiation from the clock laser, and the outcome of the excitation attempt is measured again in the fluorescence signal induced by the cooling laser. In the case of a single ion, its excitation into the upper, metastable level of the clock transition will prevent the scattering of fluorescence as long as the electron is "shelved". In the case of sympathetic cooling, a quantum-logic readout scheme can be applied to transfer the internal state of the clock ion to the common vibrational mode and to the fluorescence of the coolant ion [4].

The combination of laser cooling and ion trapping provides ideal conditions for precision spectroscopy in eliminating line broadening and frequency shifts from the Doppler effect. Already for electronic transitions, the influence of the trapping potential on transition frequencies can be very well controlled, and this carries forward even more favorably for the nuclear transition. Th$^{3+}$ has been identified as the most suitable charge state because of its simple electronic level structure with one valence electron, that provides near-infrared resonance transitions for laser cooling and fluorescence detection [5]. Direct laser cooling of $^{229}$Th$^{3+}$ is made more difficult by the hyperfine structure that requires several re-pumper laser frequencies in order to avoid hyperfine pumping into dark sublevels of the ground state. Sympathetic laser cooling by co-trapped $^{232}$Th$^{3+}$ has been demonstrated [55] and lighter ions like $^{88}$Sr$^+$ may provide a convenient alternative with similar charge-to-mass ratio. An experimental difficulty arises from the strong chemical reactivity of trapped Th ions with molecules of the residual gas. This leads to the formation of molecular ions on time scales of seconds or minutes, unless extremely clean vacuum conditions are achieved. In Th$^+$ the problem could be solved by photo dissociation with UV laser light [148], but in Th$^{3+}$ the molecule formation proceeds via charge exchange to the 2+ charge state [149], making it impossible to recover the Th$^{3+}$ ion in a single step.

The most complete elimination of field-induced frequency shifts is possible in electronic levels with selected angular momentum quantum numbers that provide either a highly symmetric electronic wave function ($J = 0$ or ½) or a decoupling of electronic and nuclear moments. This has been theoretically analyzed for Th$^{3+}$ and the two options that have been identified are the use of the metastable 7s $^2$S$_{1/2}$ state [5] or of stretched states ($F = |m_F| = I+J = 5$) of the 5f $^2$F$_{5/2}$ ground state [150]. A drawback of the first option is the limited lifetime of the state of only 0.5 s [151], that would limit the interaction time and consequently the linewidth of the nuclear laser excitation. In the stretched states, the non-zero value of $m_F$ leads to a linear Zeeman effect that has to be cancelled by averaging over symmetric Zeeman components originating from $\pm m$ states. The remaining uncertainty contributions, assumed to be controllable below $1 \cdot 10^{-18}$, are predominantly those that are due to the quadratic Doppler effect from residual motion of the ion [150], a relativistic effect that does not distinguish between electronic and nuclear transitions.

Operation of the clock would follow the protocols that are used in atomic optical clocks with single trapped ions: The ion is cooled and initialized in the ground state, the reference transition is driven with a single (Rabi) or multiple (Ramsey) pulses of the clock laser, and the outcome of the excitation attempt is probed. State detection can be performed by sequentially applying two laser beams on an electronic resonance transition, one in resonance with a hyperfine component of the nuclear ground state, the other with one of the isomer. Applying these lasers for a period during which several (e.g. >1000) fluorescence photons can be detected from the ion, would allow one to unambiguously determine the nuclear state [5]. In operation as a single-ion clock the relative frequency instability (Allan deviation) would be in the range $1 \cdot 10^{-16} (\tau/s)^{-1/2}$, for an averaging time $\tau$. Averaging the statistical uncertainty down to $1 \cdot 10^{-19}$ would require $10^6$ s. Since the Coulomb interaction between the ions has no significant effect on the nuclear resonance frequency, the $^{229}$Th$^{3+}$ nuclear clock is perfectly suited for the application of the multi-ion concept [152]. Performing the described interrogation sequence with $N$ trapped ions in parallel and with individually resolved state detection on a photon counting camera would reduce the required averaging time for reaching a given target uncertainty by a factor of $N$.

### 3.3.2 Solid-state-clock operation and performance

As already pointed out by Peik and Tamm in 2003, the weak coupling of the $^{229}$Th clock transition to external fields allows to envision a "solid-state nuclear clock", where the $^{229}$Th ions would be embedded into a transparent host material [5]. As the expected isomer excitation energy shifted into the VUV over the years (see Fig. 2.11), finding materials with suitable transparency became more challenging, converging on the family of fluoride single crystals [84,99–101].

The interaction between fields generated by a crystal lattice and the $^{229}$Th nuclear moments leads to the appearance of level shifts (homogeneous and inhomogeneous) and splittings, in analogy to the hyperfine structure [84,153]. In an ionic crystal, where all electrons are paired, fine structure is absent. These shifts and splittings are characteristic for a specific nucleus-host-crystal-system and even within a given system, several sets of spectral configurations may occur, connected to different microscopic defect configurations.

The dominating shift of the nuclear transition frequency (compared to the bare nucleus in vacuum) is the isomer shift (electric monopole shift) that arises from contact interaction between the electron cloud and the finite volume of the nucleus [154]. Extrapolating from similar systems (e.g., $^{151}$Eu$^{2+}$) studied in Mössbauer spectroscopy, the isomer shift is expected on the GHz scale for different charge-compensation configurations. The 2$^{nd}$ order Doppler effect is significantly reduced due to the tight binding of the nuclei inside the crystal matrix. At liquid nitrogen temperatures, both the 2$^{nd}$ order Doppler shift and the associated inhomogeneous broadening are well below 100 Hz.

The dominating inhomogeneous broadening mechanism of nuclear lines in a fluoride crystal emerges from the magnetic dipole interaction with fluctuating fields of the surrounding fluorine nuclei, it amounts to few kHz [153]. The nuclear quadrupole interaction with electric field gradients (EFG) leads to the emergence of a nuclear quadrupole structure as already discussed in Sec. 3.1.4. The level splittings are expected within 50-500 MHz and predominantly reflect the local EFG at the position of the nucleus.

We expect that direct laser spectroscopy of the $^{229}$Th nuclear transition will proceed over the next years (see Fig. 3.8). With progressing spectroscopic resolution, the various crystal effects will become apparent: Isomer shift at GHz resolution, quadrupole structure at MHz resolution, magnetic dipole shifts and broadenings at the kHz level.

The above-mentioned broadenings, in particular due to magnetic dipole interactions with the un-ordered surrounding nuclear moments, will cause a rapid decoherence (timescale ms) and render the commonly used Rabi/Ramsey schemes impossible in the operation of a Thorium solid-state clock. In other words, $T1$ (relaxation time) and $T2$ (coherence time) differ by many orders of magnitude. Nuclear clock operation will hence have to rely on laser stabilization via fluorescence (or absorption) spectroscopy, the achievable performance has been estimated in [153]. It crucially relies on the ratio of excitation rate $R$ to (yet unknown) decay rate $\gamma$ and hence on the available power of the interrogation laser. For $R = \gamma$, assuming $\gamma \approx 10^5$ s$^{-1}$ and $10^{14}$ interrogated nuclei, one might reach $10^{-19}$ fractional clock instability after $10^4$ s of averaging time. The estimate however neglects all possible technical constraints such as the short-term (in)stability of the interrogation laser. Note that the performance might increase if the nuclear excitation can be quenched via available electronic states, as discussed in [142], see 3.2.5.3 above.

## 4 Tests of fundamental physics and dark matter searches

### 4.1 Enhanced sensitivity of a nuclear clock to the variation of fundamental constants

If fundamental constants of nature, such as the fine-structure constant $\alpha$, change with space or time, so will atomic and nuclear energy levels, and, consequently, clock frequencies. Violation of local Lorentz invariance can also source changes in clock frequencies [155]. Many theories beyond the standard model (BSM) of particle physics lead to such variation of fundamental constants and violations of fundamental symmetries [156,157]. The gravitational potential on Earth oscillates with a period of 1 yr due to the ellipticity of Earth's orbit. Therefore, looking for variation of clock frequency ratio during the year also tests the Einstein equivalence principle (see Sec. II.I of [158]).

A variety of precision experiments with atomic clocks have been performed in search for new physics [158], looking for variations of ratios or differences of two transition frequencies over time or space, since the clocks based on different systems have different sensitivities to these effects. If technical and environmental (SM) perturbations can be excluded as their cause, any spacetime-dependent variation of clock frequency ratios would be an unambiguous signal of new physics. These studies are often on the work plan of metrology institutes as tests of the performance and reliability of newly developed clocks. In the development of optical clocks, we see the scientifically fertile situation that different systems, using a variety of elements and experimental methods are developed and are reaching competitive levels of accuracy. With different sensitivities to specific effects of new physics, meaningful comparisons can be made between a system of high sensitivity and a stable reference or "anchor" of low sensitivity. The $^{229}$Th nuclear clock would be a most interesting addition to the ensemble of clocks used in fundamental tests because it provides a strongly enhanced sensitivity for variations of fundamental constants and, consequently, for all related searches for new physics, such as searches for dark matter (DM) described in later sections.

An enhanced sensitivity of the $^{229}$Th nuclear transition frequency to changes of the coupling constants $\alpha_s$ of the strong and $\alpha$ of the electromagnetic interaction was first predicted by Flambaum [7], who pointed out that the near degeneracy of the two nuclear levels must be due to the nearly exact cancellation of $O$(MeV) differences in the contributions of different interactions to the two nuclear level energies. The sensitivity factors relating a change of the nuclear transition frequency to changes of $\alpha$ and the strong interaction parameter would be on the order of $10^5$. For comparison, the sensitivities to the $\alpha$-variation of almost all of the presently operating atomic clocks are less than 1 [159]. The highest sensitivity of an operating atomic clock to the $\alpha$-variation is −6, for an octupole transition in the Yb$^+$ ion [159]. While these sensitivities can be calculated with good accuracy for atomic clocks, this is not the case for a nuclear transition as it requires nuclear rather than atomic computations. For the dependence on $\alpha$, the sensitivity of a nuclear clock would be determined by the change in Coulomb energy $\Delta E_C$ [160]:

$1/\nu \, (d\nu/dt) = (\Delta E_C/E_I) \, (1/\alpha)(d\alpha/dt)$

Based on the reasoning that the nuclear excitation involves only the outer unpaired neutron, leaving the core proton distribution essentially unchanged, it was countered that $\Delta E_C$ could be much smaller than $O$(MeV) [160]. It was then pointed out that the Coulomb energy difference is experimentally accessible in measurements of hyperfine structure [159]. Modelling the nucleus as a uniform, hard-edged prolate ellipsoid, the change in Coulomb energy can be expressed in terms of the changes in *rms* charge radius $\langle r^2 \rangle$ and intrinsic electric quadrupole moment $Q_0$:

$\Delta E_C = -485 \, \text{MeV}((\langle r^2 \rangle^{229m}/\langle r^2 \rangle^{229})-1) + 11.6 \, \text{MeV}((Q_0^m/Q_0)-1)$

These quantities have been measured in hyperfine spectroscopy of the isomer [71] (see Sec. 2) and yield the result

$\Delta E_C = -0.29(43) \, \text{MeV}$.

With this data, a value $\Delta E_C <10$ keV cannot be excluded with certainty, but is very unlikely, so that it can be safely stated that the $\alpha$-sensitivity of the $^{229}$Th nuclear transition is significantly higher than in present optical atomic clocks. Looking separately at the contributions to $\Delta E_C$ from the *rms* radius change: $-0.16(3)$ MeV and from the change in electric quadrupole moment: $-0.13(40)$ MeV shows that an improvement in the uncertainty of the isomer's electric quadrupole moment by a factor of 20 will resolve the issue, and is within the scope of the planned experiments (see Sect 3.1.5). Recently, the $\alpha$-sensitivity $K = -0.9(3) \cdot 10^4$ has been evaluated with lower uncertainty, excluding the region $K=0$, based on the available experimental data together with nuclear models that include a relation between the change of the charge radius and that of the electric quadrupole moment [161]. The influence of the octupole deformation has also been considered and was shown to be small.

The enhanced sensitivity of the $^{229}$Th nuclear transition frequency to the value of the fine structure constant translates into an enhanced sensitivity in all searches for additional couplings of $\alpha$, like to the gravitational field or DM.

It may also be noted that the established Cs and Rb atomic clocks provide sensitivity to a nuclear property, the nuclear magnetic moment [162], but without the sensitivity enhancement that is specific for the $^{229}$Th nuclear transition. With the strong, many orders of magnitude, enhancement of the sensitivity, the nuclear clock could play the important role of a pathfinder that provides first indications of new physics.

### 4.2 Dark matter searches with a nuclear clock

Until about 5 years ago, the atomic clock investigations for the variation of the fundamental constants focused on searching for the "slow-drift" variation [105,158]. In such experiments, the ratio of the frequencies of two clocks was monitored over a period of time. As it was pointed out above, such comparisons are also needed for metrology, to ensure long-term clock stability and validate the clock uncertainty budgets. In the case of microwave clocks, such comparisons go back for over a decade [163]. These tests were initially motivated by the controversial astrophysical result that observed several sigma $\alpha$-variation in studies of quasar-absorption spectra (see Chapter II.H of [158] for a brief review), in addition to many BSM theories predicting varying constants.

It was recently shown that DM could source the variation of fundamental constants [164–166], providing new, very strong motivation for clock-comparison experiments. There is overwhelming observational evidence for the apparent existence of DM that is observed only via its gravitational interactions [167]. This mystery dates back to the 1930s [168] and is confirmed by numerous studies of astronomical objects at various scales, which show that the particles of the SM make up only about 16% of the total matter present in our Universe [169]. Despite decades of investigations, we have not identified the nature of dark matter. The difficulty in directly detecting the dark matter lies in the vast parameter space – we neither know the mass of the dark matter particles nor the strength of their non-gravitational interaction with the SM. However, unlike some other BSM searches, we have a definitive quantity that we do actually know about this "new physics" – a dark matter density in our Galaxy is $\rho_{DM} \approx 0.3$-$0.4\,\text{GeV/cm}^3$, corresponding to one hydrogen atom per a few cubic cm. A huge dark matter halo extends far outside of the visible galactic disk [170]. This is why the total dark matter mass in our galaxy so much exceeds the "normal matter" mass. Note that the actual DM mass in the Solar system is small, explaining why we so far did not see any DM effect on the motion of objects in a solar system. Lunar ranging experiments set a limit on a maximum dark matter mass within the moon orbit [171].

The mass of the dark matter particles is unknown, and it is bound by about $10^{-22}\,\text{eV}$ on the light side [172] and by the Planck mass on the heavy side (anything heavier would be a black hole). The "light" side limit assumes that a single particle type is responsible for a large fraction of DM and can be lowered in certain models. Dark matter with mass less than $\sim 1\,\text{eV}$ is generally referred to as "ultralight". It has to be bosonic since the Fermi velocity for fermionic DM with mass <10 eV is higher than our galaxy escape velocity [173]. Such light scalar fields arise in many BSM theories.

Various DM scenarios have been proposed that are potentially detectable with clocks. We note that clocks are sensitive to a broad range of ultralight scalar particles, which do not necessarily have to be a significant DM component, avoiding the lower mass limit above.

### 4.2.1 Oscillatory signals

Light DM particles have large mode occupation numbers, and their phenomenology is described by a classical field [165]. Optical clocks and other precision measurement techniques are well suited for detecting such dark matter candidates that act as coherent entities on the scale of individual detectors or networks of detectors. Optical clocks are sensitive to ultralight scalar DM that can be described as a field oscillating at the Compton frequency of the dark matter particles [165]. The coupling of such an oscillating field to the SM leads to an oscillation of fundamental constants, such as the fine-structure constant $\alpha$ and, therefore, atomic and nuclear energies. These result in the oscillation of clock frequencies leading to persistent time-varying signals that may be detected by monitoring ratios of clock frequencies. Previous searches for the slow-drift $\alpha$-variation with Dy and Rb/Cs clocks were re-analyzed in terms of dark matter limits in [174,175]. A method to search for fast oscillations of fundamental constants using atomic spectroscopy in cesium vapor was reported in [176], demonstrating sensitivities to higher DM masses in the range from $8\cdot10^{-11}$ to $4\cdot10^{-7}\,\text{eV}$.

The frequency of a resonant mode of an ultrastable crystalline silicon optical cavity is also sensitive to the variation of the fundamental constants [166]. This sensitivity is directly traceable to the Bohr radius: the cavity spacer length is proportional to the silicon crystal lattice constant [177]. Therefore, such ultrastable cavities are also suitable for the clock-cavity DM searches. New frequency comparisons between a state-of-the-art strontium optical lattice clock, a cryogenic crystalline silicon cavity, and a hydrogen maser set new bounds on the coupling of ultralight dark matter to SM particles and fields in the mass range of $10^{-16}$-$10^{-21}\,\text{eV}$ [177].

Very high sensitivity of a nuclear clock to the variation of $\alpha$ directly translates into a high sensitivity to oscillating effects as well as transient effects sourced by ultralight scalar dark matter. Therefore, the nuclear clock is expected to probe DM with sensitivity potentially exceeding present limits by many orders of magnitude. Moreover, the nuclear clock is also sensitive to the changes in fundamental constants other than $\alpha$, i.e., related to the strong interaction. This means that a nuclear clock is sensitive to DM coupling to the nuclear rather than just to the electromagnetic sector of the SM [7], leading to sensitivities to different DM-SM couplings, increasing the discovery potential further.

### 4.2.2 Transient signals

If the dark matter contains some self-interaction, it may form potentially large-scale clumps (such as topological defects), leading to transient changes in the fundamental constants [164,166].

If the DM fields have non-gravitational interactions with SM fields, encounters between such objects and precision measurement devices may induce observable transient signatures in clock-comparison and clock-cavity comparison data as the Earth moves through the galactic dark matter halo [15].

Such DM can manifest, for example, as "glitches" in atomic clock networks, either in space onboard satellites of the Global Positioning System (GPS) clocks or on Earth. Both GPS [178] and ground-based clocks [179–181] have been used for such searches. As a network of clocks is swept by such extended (potentially Earth-size) DM objects at galactic velocities, the clocks will become desynchronized due to transient DM, but then will recover syntonization. The characteristic velocity here is the velocity with which Sun orbits the center of the Milky Way Galaxy and, therefore, moves through the dark matter halo, at about 230 km/s. A network of clocks (as opposed to a single clock) distinguishes a transient frequency variation due to DM from one caused by terrestrial sources. Specifically, time delays between signals appearing across the clock network must be consistent with the passing of a DM transient object, taking about a minute to sweep the Earth diameter. Moreover, the use of clocks with different sensitivities to the $\alpha$-variation allows additional rejection of false positives based on known sensitivities of different clocks. New limits on such transient DM have been recently reported by [181] using a European network of fiber-linked optical atomic clocks. Ref. [181] considered the time-scales of over 60 s, where transient DM would manifest as a simultaneous signal visible in all data streams.

A network of three nuclear clocks, operating even at $10^{-15}$ precision at three locations, such as LMU Munich, PTB Braunschweig, and TU Vienna, will allow to significantly improve current limits due to the orders of magnitude higher sensitivity. If the projected $10^{-19}$ accuracy [150] of nuclear clocks is achieved, such a network will become an exceptional tool in detecting both oscillatory and transient DM signals.

*4.2.3 Hierarchy problem, relaxion, and a nuclear clock*

Aside from our lack of understanding of the nature of DM, the SM of fundamental interactions and elementary particles fails to answer a number of other fundamental questions. For example, there is a matter-antimatter asymmetry problem: a glaring lack of primordial (i.e., produced shortly after the Big Bang) antimatter in the present-day Universe. Another is a hierarchy (or a fine-tuning or a naturalness) problem. The hierarchy problem is related to the question as to why the weak force is so much stronger than the gravitational force. Within the SM, the Higgs boson mass is ultra-sensitive to any form of heavy new degrees of freedom. Therefore, the large quantum contributions to the square of the Higgs boson mass generically push it to the largest scale in the problem (for example, Planck mass) unless these are (extremely precise) fine-tuning cancellations [182].

A number of solutions are proposed to this problem, with variants of supersymmetry being particularly attractive as the lightest supersymmetric particle also provide a dark matter candidate that falls into a category of a "weakly interactive massive particle", i.e. a ~TeV-scale WIMP [183]. Lack of either observation of supersymmetric particles at the Large Hadron Collider (LHC) or a generic WIMPs over decades of experiments motivated searches for alternative solutions.

In 2015, a dynamical relaxation was proposed to account for the lightness of the Higgs without invoking new particles at the TeV scale [184]. The relaxion framework involves an additional new physics field in the form of an ultralight scalar, relaxion, which dynamically relaxes the Higgs mass with respect to its natural large value [185]. The relaxion is not accessible to the LHC or other energy-frontier detectors. In addition, the relaxion makes a viable light dark matter candidate [186] and can address the baryon asymmetry of the Universe [187]. A requirement that the relaxion solves a hierarchy problem produces constraints on the strength of its interaction with the SM particles. Therefore, from the standpoint of the ultralight particle searches with clocks, the relaxion presents a particularly interesting candidate with a well-defined parameter space and concrete benchmarks that we have to aim for. The parameter space for the light relaxion and a projected reach of the nuclear clock was recently studied in [188]. The ability of the nuclear clock to reach the relaxion parameter space is based on its unique sensitivity to the coupling of new physics to the nuclear sector of the SM and its large sensitivity.

## 5 Conclusion

In this paper we described a route towards a nuclear clock focusing on the planned efforts of the teams forming the ERC Synergy Grant 'ThoriumNuclearClock' consortium. We described the current knowledge on the properties of the $^{229m}$Th nuclear clock transition and planned efforts to advance this knowledge, with different VUV laser sources for driving the transition, possibly assisted by a bridge process in the electron shell. We sketched the strategy towards the realization of a thorium nuclear clock with different experimental schemes, using thorium ions in a Paul trap or implanted in a VUV-transparent crystal. Finally, we outlined the potential of the nuclear clock for testing fundamental physics, especially the search for time variation of fundamental constants and the search for dark matter.

## Acknowledgements

This work is part of the 'ThoriumNuclearClock' project that has received funding from the European Research Council (ERC) under the European Union's Horizon 2020 research and innovation program (Grant Agreement No. 856415). Work at PTB and Vienna University of Technology is supported by the project EMPIR 17FUN07 "Coulomb Crystals for Clocks". This project has received funding from the EMPIR programme co-financed by the Participating States and from the European Union's Horizon 2020 research and innovation programme. The Vienna team acknowledges support by the Austrian Science Fund (FWF): P33627. Fruitful discussions with Gilad Perez and Abhishek Banerjee are gratefully acknowledged. We acknowledge technical support by Tomas

## References

[1] W. P. Schleich, K. S. Ranade, C. Anton, M. Arndt, M. Aspelmeyer, M. Bayer, G. Berg, T. Calarco, H. Fuchs, E. Giacobino, M. Grassl, P. Hänggi, W. M. Heckl, I.-V. Hertel, S. Huelga, F. Jelezko, B. Keimer, J. P. Kotthaus, G. Leuchs, N. Lütkenhaus, U. Maurer, T. Pfau, M. B. Plenio, E. M. Rasel, O. Renn, C. Silberhorn, J. Schiedmayer, D. Schmitt-Landsiedel, K. Schönhammer, A. Ustinov, P. Walther, H. Weinfurter, E. Welzl, R. Wiesendanger, S. Wolf, A. Zeilinger, and P. Zoller, Appl. Phys. B **122**, 130 (2016).
[2] T. Quinn, in *Metrologia*, Vol. 42 (2005).
[3] A. D. Ludlow, M. M. Boyd, J. Ye, E. Peik, and P. O. Schmidt, Rev. Mod. Phys. **87**, 637 (2015).
[4] S. M. Brewer, J.-S. Chen, A. M. Hankin, E. R. Clements, C. W. Chou, D. J. Wineland, D. B. Hume, and D. R. Leibrandt, Phys. Rev. Lett. **123**, 033201 (2019).
[5] E. Peik and C. Tamm, Europhys. Lett. **61**, 181 (2003).
[6] E. Peik and M. Okhapkin, Comptes Rendus Physique.
[7] V. V Flambaum, Phys. Rev. Lett. **97**, 92502 (2006).
[8] B. N. L. National Nuclear Data Center, National Nuclear Data Center; http://www.nndc.bnl.gov.
[9] L. A. Kroger and C. W. Reich, Nucl. Physics, Sect. A **259**, 29 (1976).
[10] C. W. Reich and R. G. Helmer, Phys. Rev. Lett. **64**, 271 (1990).
[11] D. G. Burke, P. E. Garrett, T. Qu, and R. A. Naumann, Phys. Rev. C **42**, R499 (1990).
[12] R. G. Helmer and C. W. Reich, Phys. Rev. C **49**, 1845 (1994).
[13] O. V Vorykhalov and V. V Koltsov, Bull. Rus. Acad. Sci. Phys. **59**, 20 (1995).
[14] K. Gulda, W. Kurcewicz, A. J. Aas, M. J. G. Borge, D. G. Burke, B. Fogelberg, I. S. Grant, E. Hagebo, N. Kaffrell, J. Kvasil, G. Lovhoiden, H. Mach, A. Mackova, T. Martinez, G. Nyman, B. Rubio, J. L. Tain, O. Tengblad, and T. F. Thorsteinsen, Nucl. Phys. A **703**, 45 (2002).
[15] V. Barci, G. Ardisson, G. Barci-Funel, B. Weiss, O. El Samad, and R. K. Sheline, Phys. Rev. C - Nucl. Phys. **68**, 23 (2003).
[16] I. D. Moore, I. Ahmad, K. Bailey, D. L. Bowers, Z. T. Lu, T. P. O'Connor, and Z. Yin, Argonne Phys. Div. Rep. PHY-10990-ME-2004 (2004).
[17] Z. O. Guimarães-Filho and O. Helene, Phys. Rev. C - Nucl. Phys. **71**, 44303 (2005).
[18] E. Ruchowska, W. A. Płóciennik, J. Zylicz, H. Mach, J. Kvasil, A. Algora, N. Amzal, T. Bäck, M. G. Borge, R. Boutami, P. A. Butler, J. Cederkäll, B. Cederwall, B. Fogelberg, L. M. Fraile, H. O. U. Fynbo, E. Hagebø, P. Hoff, H. Gausemel, A. Jungclaus, R. Kaczarowski, A. Kerek, W. Kurcewicz, K. Lagergren, E. Nacher, B. Rubio, A. Syntfeld, O. Tengblad, A. A. Wasilewski, and L. Weissman, Phys. Rev. C - Nucl. Phys. **73**, 44326 (2006).
[19] LINSTROM, P. J. Ed., and P. J. Linstrom, NIST Chem. Webb. (2003).
[20] G. M. Irwin and K. H. Kim, Phys. Rev. Lett. **79**, 990 (1997).
[21] D. S. Richardson, D. M. Benton, D. E. Evans, J. A. R. Griffith, and G. Tungate, Phys. Rev. Lett. **80**, 3206 (1998).
[22] R. W. Shaw, J. P. Young, S. P. Cooper, and O. F. Webb, Phys. Rev. Lett. **82**, 1109 (1999).
[23] S. B. Utter, P. Beiersdorfer, A. Barnes, R. W. Lougheed, J. R. Crespo López-Urrutia, J. A. Becker, and M. S. Weiss, Phys. Rev. Lett. **82**, 505 (1999).
[24] J. P. Young, R. W. Shaw, and O. F. Webb, Inorg. Chem. **38**, 5192 (1999).
[25] T. Mitsugashira, M. Hara, T. Ohtsuki, H. Yuki, K. Takamiya, Y. Kasamatsu, A. Shinohara, H. Kikunaga, and T. Nakanishi, J. Radioanal. Nucl. Chem. **255**, 63 (2003).
[26] E. Browne, E. B. Norman, R. D. Canaan, D. C. Glasgow, J. M. Keller, and J. P. Young, Phys. Rev. C - Nucl. Phys. **64**, 143111 (2001).
[27] Y. Kasamatsu, H. Kikunaga, K. Takamiya, T. Mitsugashira, T. Nakanishi, Y. Ohkubo, T. Ohtsuki, W. Sato, and A. Shinohara, in *Radiochimica Acta*, Vol. 93 (De Gruyter, 2005), pp. 511–514.
[28] H. Kikunaga, Y. Kasamatsu, K. Takamiya, T. Mitsugashira, M. Hara, T. Ohtsuki, H. Yuki, A. Shinohara, S. Shibata, N. Kinoshita, A. Yokoyama, and T. Nakanishi, in *Radiochimica Acta*, Vol. 93 (De Gruyter, 2005), pp. 507–510.
[29] E. V Tkalya, JETP Lett. **70**, 371 (1999).
[30] E. V Tkalya, A. N. Zherikhin, and V. I. Zhudov, Phys. Rev. C - Nucl. Phys. **61**, 6 (2000).
[31] F. F. Karpeshin and M. B. Trzhaskovskaya, Phys. Rev. C - Nucl. Phys. **76**, 54313 (2007).
[32] E. V Tkalya, V. O. Varlamov, V. V Lomonosov, and S. A. Nikulin, Phys. Scr. **53**, 296 (1996).
[33] F. F. Karpeshin, I. M. Band, M. B. Trzhaskovskaya, and M. A. Listengarten, Phys. Lett. Sect. B Nucl. Elem. Part. High-Energy Phys. **372**, 1 (1996).
[34] A. M. Dykhne and E. V Tkalya, JETP Lett. **67**, 549 (1998).
[35] F. F. Karpeshin, I. M. Band, and M. B. Trzhaskovskaya, Nucl. Phys. A **654**, 579 (1999).
[36] T. T. Inamura and T. Mitsugashira, in *Laser 2004* (Springer-Verlag, 2006), pp. 115–123.
[37] F. F. Karpeshin and M. B. Trzhaskovskaya, Yad. Fiz. **69**, 596 (2006).
[38] S. Matinyan, Phys. Rep. **298**, 199 (1998).
[39] E. V. Tkalya, Physics-Uspekhi **46**, 315 (2003).
[40] B. R. Beck, J. A. Becker, P. Beiersdorfer, G. V Brown, K. J. Moody, J. B. Wilhelmy, F. S. Porter, C. A. Kilbourne, and R. L. Kelley, Phys. Rev. Lett. **98**, 142501 (2007).
[41] B. R. Beck, C. Wu, P. Beiersdorfer, G. V Brown, J. A. Becker, K. J. Moody, J. B. Wilhelmy, F. S. Porter, C. A. Kilbourne, and R. L. Kelley, LLNL-PROC-415170 TRN US0903866 (2009).
[42] X. Zhao, Y. N. Martinez De Escobar, R. Rundberg, E. M. Bond, A. Moody, and D. J. Vieira, Phys. Rev. Lett. **109**, 160801 (2012).
[43] E. Peik and K. Zimmermann, Phys. Rev. Lett. **111**, 18901 (2013).
[44] M. P. Hehlen, R. R. Greco, W. G. Rellergert, S. T. Sullivan, D. Demille, R. A. Jackson, E. R. Hudson, and J. R. Torgerson, in *Journal of Luminescence*, Vol. 133 (North-

[45] J. Jeet, C. Schneider, S. T. Sullivan, W. G. Rellergert, S. Mirzadeh, A. Cassanho, H. P. Jenssen, E. V Tkalya, and E. R. Hudson, Phys. Rev. Lett. **114**, 253001 (2015).

[46] S. Stellmer, G. Kazakov, M. Schreitl, H. Kaser, M. Kolbe, and T. Schumm, Phys. Rev. A **97**, 62506 (2018).

[47] S. Stellmer, M. Schreitl, G. A. Kazakov, J. H. Sterba, and T. Schumm, Phys. Rev. C **94**, 14302 (2016).

[48] A. Yamaguchi, M. Kolbe, H. Kaser, T. Reichel, A. Gottwald, and E. Peik, New J. Phys. **17**, 53053 (2015).

[49] L. v. d. Wense, P. G. Thirolf, D. Kalb, and M. Laatiaoui, J. Instrum. **8**, P03005 (2013).

[50] K. Zimmermann, No Title, University of Hannover, 2010.

[51] S. G. Porsev, V. V Flambaum, E. Peik, and C. Tamm, Phys. Rev. Lett. **105**, 182501 (2010).

[52] K. Beloy, Phys. Rev. Lett. **112**, 62503 (2014).

[53] P. V Bilous and L. P. Yatsenko, Ukr. J. Phys. **60**, 371 (2015).

[54] F. F. Karpeshin and M. B. Trzhaskovskaya, Phys. At. Nucl. **78**, 715 (2015).

[55] C. J. Campbell, A. G. Radnaev, and A. Kuzmich, Phys. Rev. Lett. **106**, 223001 (2011).

[56] V. Sonnenschein, I. D. Moore, S. Raeder, A. Hakimi, A. Popov, and K. Wendt, in *Three Decades of Research Using IGISOL Technique at the University of Jyväskylä* (Springer Netherlands, 2012), pp. 311–325.

[57] E. V Tkalya, C. Schneider, J. Jeet, and E. R. Hudson, Phys. Rev. C - Nucl. Phys. **92**, 54324 (2015).

[58] P. G. Thirolf, B. Seiferle, and L. v. d. Wense, Ann. Phys. **531**, 1800381 (2019).

[59] P. G. Thirolf, B. Seiferle, and L. v. d. Wense, J. Phys. B At. Mol. Opt. Phys. **52**, 203001 (2019).

[60] L. v. d. Wense, B. Seiferle, and P. G. Thirolf, Meas. Tech. **60**, 1178 (2018).

[61] L. v. d. Wense, B. Seiferle, M. Laatiaoui, J. B. Neumayr, H. J. Maier, H. F. Wirth, C. Mokry, J. Runke, K. Eberhardt, C. E. Düllmann, N. G. Trautmann, and P. G. Thirolf, Nature **533**, 47 (2016).

[62] https://cordis.europa.eu/project/id/856415/de.

[63] B. Seiferle, L. v. d. Wense, P. V. Bilous, I. Amersdorffer, C. Lemell, F. Libisch, S. Stellmer, T. Schumm, C. E. Düllmann, A. Pálffy, and P. G. Thirolf, Nature **573**, 243 (2019).

[64] T. Sikorsky, J. Geist, D. Hengstler, S. Kempf, L. Gastaldo, C. Enss, C. Mokry, J. Runke, C. E. Düllmann, P. Wobrauschek, K. Beeks, V. Rosecker, J. H. Sterba, G. Kazakov, T. Schumm, and A. Fleischmann, Phys. Rev. Lett. **125**, 142503 (2020).

[65] J. B. Neumayr, P. G. Thirolf, D. Habs, S. Heinz, V. S. Kolhinen, M. Sewtz, and J. Szerypo, Rev. Sci. Instrum. **77**, 65109 (2006).

[66] L. v. d. Wense, B. Seiferle, M. Laatiaoui, and P. G. Thirolf, Eur. Phys. J. A **51**, 29 (2015).

[67] L. v. d. Wense, B. Seiferle, I. Amersdorffer, and P. G. Thirolf, J. Vis. Exp. **2019**, 58516 (2019).

[68] B. Seiferle, L. v. d. Wense, and P. G. Thirolf, Phys. Rev. Lett. **118**, 42501 (2017).

[69] B. Seiferle, No Title, LMU Munich, 2019.

[70] F. F. Karpeshin and M. B. Trzhaskovskaya, Nucl. Phys. A **969**, 173 (2018).

[71] J. Thielking, M. V Okhapkin, P. Głowacki, D. M. Meier, L. v. d. Wense, B. Seiferle, C. E. Düllmann, P. G. Thirolf, and E. Peik, Nature **556**, 321 (2018).

[72] N. Minkov and A. Pálffy, Phys. Rev. Lett. **122**, 162502 (2019).

[73] M. S. Safronova, S. G. Porsev, M. G. Kozlov, J. Thielking, M. V. Okhapkin, P. Głowacki, D. M. Meier, and E. Peik, Phys. Rev. Lett. **121**, 213001 (2018).

[74] B. Seiferle, L. v. d. Wense, I. Amersdorffer, N. Arlt, B. Kotulski, and P. G. Thirolf, Nucl. Instruments Methods Phys. Res. Sect. B Beam Interact. with Mater. Atoms **463**, 499 (2020).

[75] R. G. Helmer and C. W. Reich, Int. J. Appl. Radiat. Isot. **36**, 117 (1985).

[76] G. A. Kazakov, T. Schumm, and S. Stellmer, Http://Arxiv.Org/Abs/1702.00749 (2017).

[77] https://zenodo.org/communities/thorium_nuclear_clock/search?page=1&size=20.

[78] A. Yamaguchi, H. Muramatsu, T. Hayashi, N. Yuasa, K. Nakamura, M. Takimoto, H. Haba, K. Konashi, M. Watanabe, H. Kikunaga, K. Maehata, N. Y. Yamasaki, and K. Mitsuda, Phys. Rev. Lett. **123**, 222501 (2019).

[79] T. Masuda, A. Yoshimi, A. Fujieda, H. Fujimoto, H. Haba, H. Hara, T. Hiraki, H. Kaino, Y. Kasamatsu, S. Kitao, K. Konashi, Y. Miyamoto, K. Okai, S. Okubo, N. Sasao, M. Seto, T. Schumm, Y. Shigekawa, K. Suzuki, S. Stellmer, K. Tamasaku, S. Uetake, M. Watanabe, T. Watanabe, Y. Yasuda, A. Yamaguchi, Y. Yoda, T. Yokokita, M. Yoshimura, and K. Yoshimura, Nature **573**, 238 (2019).

[80] M. Seto, J. Phys. Soc. Japan **82**, (2013).

[81] A. Yoshimi, H. Hara, T. Hiraki, Y. Kasamatsu, S. Kitao, Y. Kobayashi, K. Konashi, R. Masuda, T. Masuda, Y. Miyamoto, K. Okai, S. Okubo, R. Ozaki, N. Sasao, O. Sato, M. Seto, T. Schumm, Y. Shigekawa, S. Stellmer, K. Suzuki, S. Uetake, M. Watanabe, A. Yamaguchi, Y. Yasuda, Y. Yoda, K. Yoshimura, and M. Yoshimura, Phys. Rev. C **97**, 24607 (2018).

[82] T. Masuda, T. Watanabe, K. Beeks, H. Fujimoto, T. Hiraki, H. Kaino, S. Kitao, Y. Miyamoto, K. Okai, N. Sasao, M. Seto, T. Schumm, Y. Shigekawa, K. Tamasaku, S. Uetake, A. Yamaguchi, Y. Yoda, A. Yoshimi, and K. Yoshimura, Absolute X-Ray Energy Measurement Using a High-Accuracy Angle Encoder, 2020.

[83] W. L. Bond, Acta Crystallogr. **13**, 814 (1960).

[84] W. G. Rellergert, D. DeMille, R. R. Greco, M. P. Hehlen, J. R. Torgerson, and E. R. Hudson, Phys. Rev. Lett. **104**, 200802 (2010).

[85] G. Porat, C. M. Heyl, S. B. Schoun, C. Benko, N. Dörre, K. L. Corwin, and J. Ye, in *Nonlinear Optics (NLO)* (OSA, Washington, D.C., 2019).

[86] L. v. d. Wense and C. Zhang, Eur. Phys. J. D **74**, 146 (2020).

[87] P. V Borisyuk, S. P. Derevyashkin, K. Yu Khabarova, N. N. Kolachevsky, Y. Y. Lebedinsky, S. S. Poteshin, A. A. Sysoev, E. V Tkalya, D. O. Tregubov, V. I. Troyan, O. S. Vasiliev, and V. P. Yakovlev, J. Phys. Conf. Ser. **941**, 012107 (2017).

[88] P. V. Borisyuk, N. N. Kolachevsky, A. V. Taichenachev, E.

V. Tkalya, I. Y. Tolstikhina, and V. I. Yudin, Phys. Rev. C **100**, 044306 (2019).

[89] https://www.opticlock.de/.

[90] B. Seiferle, L. v. d. Wense, and P. G. Thirolf, Eur. Phys. J. A **53**, 1 (2017).

[91] M. Verlinde, S. Kraemer, J. Moens, K. Chrysalidis, J. G. Correia, S. Cottenier, H. De Witte, D. V Fedorov, V. N. Fedosseev, R. Ferrer, L. M. Fraile, S. Geldhof, C. A. Granados, M. Laatiaoui, T. A. L. Lima, P. C. Lin, V. Manea, B. A. Marsh, I. Moore, L. M. C. Pereira, S. Raeder, P. Van Den Bergh, P. Van Duppen, A. Vantomme, E. Verstraelen, U. Wahl, and S. G. Wilkins, Phys. Rev. C **100**, 24315 (2019).

[92] O. Kofoed-Hansen, Phys. Rev. **74**, 1785 (1948).

[93] M. N. De Mévergnies, Phys. Rev. Lett. **23**, 422 (1969).

[94] M. N. De Mevergnies, Phys. Rev. Lett. **29**, 1188 (1972).

[95] M. Schwarz, O. O. Versolato, A. Windberger, F. R. Brunner, T. Ballance, S. N. Eberle, J. Ullrich, P. O. Schmidt, A. K. Hansen, A. D. Gingell, M. Drewsen, and J. R. C. López-Urrutia, Rev. Sci. Instrum. **83**, 83115 (2012).

[96] L. Schmöger, O. O. Versolato, M. Schwarz, M. Kohnen, A. Windberger, B. Piest, S. Feuchtenbeiner, J. Pedregosa-Gutierrez, T. Leopold, P. Micke, A. K. Hansen, T. M. Baumann, M. Drewsen, J. Ullrich, P. O. Schmidt, and J. R. C. López-Urrutia, Science (80-. ). **347**, 1233 (2015).

[97] D. Moritz, No Title, LMU Munich, 2020.

[98] M. S. Safronova, U. I. Safronova, A. G. Radnaev, C. J. Campbell, and A. Kuzmich, Phys. Rev. A **88**, 60501 (2013).

[99] R. A. Jackson, J. B. Amaral, M. E. G. Valerio, D. P. DeMille, and E. R. Hudson, J. Phys. Condens. Matter **21**, 325403 (2009).

[100] P. Dessovic, P. Mohn, R. A. Jackson, G. Winkler, M. Schreitl, G. Kazakov, and T. Schumm, J. Phys. Condens. Matter **26**, 105402 (2014).

[101] M. Pimon, J. Gugler, P. Mohn, G. A. Kazakov, N. Mauser, and T. Schumm, J. Phys. Condens. Matter **32**, 255503 (2020).

[102] M. S. Safronova and W. R. Johnson, *All-Order Methods for Relativistic Atomic Structure Calculations* (2008), pp. 191–233.

[103] M. Fan, C. A. Holliman, S. G. Porsev, M. S. Safronova, and A. M. Jayich, Phys. Rev. A **100**, 062504 (2019).

[104] M. S. Safronova, U. I. Safronova, and C. W. Clark, Phys. Rev. A **90**, 032512 (2014).

[105] M. S. Safronova, Ann. Phys. **531**, 1800364 (2019).

[106] M. V. Okhapkin, D. M. Meier, E. Peik, M. S. Safronova, M. G. Kozlov, and S. G. Porsev, Phys. Rev. A **92**, 020503 (2015).

[107] C. Cheung, M. S. Safronova, S. G. Porsev, M. G. Kozlov, I. I. Tupitsyn, and A. I. Bondarev, Phys. Rev. Lett. **124**, 163001 (2020).

[108] M. S. Safronova, C. Cheung, M. G. Kozlov, S. E. Spielman, N. D. Gibson, and C. W. Walter, (2020).

[109] L. v. d. Wense, B. Seiferle, S. Stellmer, J. Weitenberg, G. Kazakov, A. Pálffy, and P. G. Thirolf, Phys. Rev. Lett. **119**, 132503 (2017).

[110] S. J. Hanna, P. Campuzano-Jost, E. A. Simpson, D. B. Robb, I. Burak, M. W. Blades, J. W. Hepburn, and A. K. Bertram, Int. J. Mass Spectrom. **279**, 134 (2009).

[111] A. Comby, D. Descamps, S. Beauvarlet, A. Gonzalez, F. Guichard, S. Petit, Y. Zaouter, and Y. Mairesse, Opt. Express **27**, (2019).

[112] P. Russbueldt, D. Hoffmann, M. Hofer, J. Lohring, J. Luttmann, A. Meissner, J. Weitenberg, M. Traub, T. Sartorius, D. Esser, R. Wester, P. Loosen, and R. Poprawe, IEEE J. Sel. Top. Quantum Electron. **21**, (2015).

[113] A. Ruehl, A. Marcinkevicius, M. E. Fermann, and I. Hartl, Opt. Lett. **35**, (2010).

[114] N. Lilienfein, C. Hofer, S. Holzberger, C. Matzer, P. Zimmermann, M. Trubetskov, V. Pervak, and I. Pupeza, Opt. Lett. **42**, (2017).

[115] J. Schulte, T. Sartorius, J. Weitenberg, A. Vernaleken, and P. Russbueldt, Opt. Lett. **41**, (2016).

[116] J. Weitenberg, T. Saule, J. Schulte, and P. Rusbuldt, IEEE J. Quantum Electron. **53**, (2017).

[117] T. Saule, S. Heinrich, J. Schötz, N. Lilienfein, M. Högner, O. deVries, M. Plötner, J. Weitenberg, D. Esser, J. Schulte, P. Russbueldt, J. Limpert, M. F. Kling, U. Kleineberg, and I. Pupeza, Nat. Commun. **10**, (2019).

[118] A. K. Mills, S. Zhdanovich, M. X. Na, F. Boschini, E. Razzoli, M. Michiardi, A. Sheyerman, M. Schneider, T. J. Hammond, V. Süss, C. Felser, A. Damascelli, and D. J. Jones, Rev. Sci. Instrum. **90**, (2019).

[119] A. Ozawa, Z. Zhao, M. Kuwata-Gonokami, and Y. Kobayashi, Opt. Express **23**, (2015).

[120] C. Zhang, S. B. Schoun, C. M. Heyl, G. Porat, M. B. Gaarde, and J. Ye, Phys. Rev. Lett. **125**, (2020).

[121] C. Benko, T. K. Allison, A. Cingöz, L. Hua, F. Labaye, D. C. Yost, and J. Ye, Nat. Photonics **8**, (2014).

[122] A. Cingöz, D. C. Yost, T. K. Allison, A. Ruehl, M. E. Fermann, I. Hartl, and J. Ye, Nature **482**, (2012).

[123] A. Ozawa and Y. Kobayashi, Phys. Rev. A **87**, (2013).

[124] T. T. Luu, M. Garg, S. Y. Kruchinin, A. Moulet, M. T. Hassan, and E. Goulielmakis, Nature **521**, 498 (2015).

[125] G. Ndabashimiye, S. Ghimire, M. Wu, D. A. Browne, K. J. Schafer, M. B. Gaarde, and D. A. Reis, Nature **534**, 520 (2016).

[126] A. A. Lanin, E. A. Stepanov, A. B. Fedotov, and A. M. Zheltikov, Optica **4**, 516 (2017).

[127] Y. S. You, M. Wu, Y. Yin, A. Chew, X. Ren, S. Gholam-Mirzaei, D. A. Browne, M. Chini, Z. Chang, K. J. Schafer, M. B. Gaarde, and S. Ghimire, Opt. Lett. **42**, 1816 (2017).

[128] J. Seres, E. Seres, C. Serrat, and T. Schumm, Opt. Express **26**, 21900 (2018).

[129] J. Seres, E. Seres, C. Serrat, E. C. Young, J. S. Speck, and T. Schumm, Opt. Express **27**, 6618 (2019).

[130] K. Wakui, K. Hayasaka, and T. Ido, Appl. Phys. B **117**, 957 (2014).

[131] C. Gohle, T. Udem, M. Herrmann, J. Rauschenberger, R. Holzwarth, H. A. Schuessler, F. Krausz, and T. W. Hänsch, Nature **436**, 234 (2005).

[132] V. A. Krutov and K. Müller, Bull. Acad. Sci. USSR (Phys. Ser) **22**, 171 (1958).

[133] V. A. Krutov and K. Müller, Bull. Acad. Sci. USSR (Phys. Ser) **22**, 161 (1958).

[134] V. F. Strizhof and E. V Tkalya, Sov. Phys.J ETP **72**, 387

(1991).

[135] E. V Tkalya, Sov. Phys. JETP **75**, 200 (1992).

[136] S. G. Porsev and V. V. Flambaum, Phys. Rev. A **81**, 032504 (2010).

[137] S. G. Porsev and V. V. Flambaum, Phys. Rev. A **81**, 042516 (2010).

[138] P. V. Bilous, Towards a Nuclear Clock with the 229Th Isomeric Transition, Universität Heidelberg, 2018.

[139] D.-M. Meier, J. Thielking, P. Głowacki, M. V. Okhapkin, R. A. Müller, A. Surzhykov, and E. Peik, Phys. Rev. A **99**, 052514 (2019).

[140] P. V Bilous, E. Peik, and A. Pálffy, New J. Phys. **20**, 013016 (2018).

[141] P. V. Bilous, H. Bekker, J. C. Berengut, B. Seiferle, L. v. d. Wense, P. G. Thirolf, T. Pfeifer, J. R. C. López-Urrutia, and A. Pálffy, Phys. Rev. Lett. **124**, 192502 (2020).

[142] B. S. Nickerson, M. Pimon, P. V. Bilous, J. Gugler, K. Beeks, T. Sikorsky, P. Mohn, T. Schumm, and A. Pálffy, Phys. Rev. Lett. **125**, 032501 (2020).

[143] K. Bergmann, H.-C. Nägerl, C. Panda, G. Gabrielse, E. Miloglyadov, M. Quack, G. Seyfang, G. Wichmann, S. Ospelkaus, A. Kuhn, S. Longhi, A. Szameit, P. Pirro, B. Hillebrands, X.-F. Zhu, J. Zhu, M. Drewsen, W. K. Hensinger, S. Weidt, T. Halfmann, H.-L. Wang, G. S. Paraoanu, N. V Vitanov, J. Mompart, T. Busch, T. J. Barnum, D. D. Grimes, R. W. Field, M. G. Raizen, E. Narevicius, M. Auzinsh, D. Budker, A. Pálffy, and C. H. Keitel, J. Phys. B At. Mol. Opt. Phys. **52**, 202001 (2019).

[144] J. C. Berengut, Phys. Rev. Lett. **121**, 253002 (2018).

[145] V. Mäckel, R. Klawitter, G. Brenner, J. R. Crespo López-Urrutia, and J. Ullrich, Phys. Rev. Lett. **107**, 143002 (2011).

[146] G. W. Rubloff, Phys. Rev. B **5**, 662 (1972).

[147] G. Kresse and D. Joubert, Phys. Rev. B **59**, 1758 (1999).

[148] O. A. Herrera-Sancho, N. Nemitz, M. V. Okhapkin, and E. Peik, Phys. Rev. A **88**, 012512 (2013).

[149] L. R. Churchill, M. V. DePalatis, and M. S. Chapman, Phys. Rev. A **83**, 012710 (2011).

[150] C. J. Campbell, A. G. Radnaev, A. Kuzmich, V. A. Dzuba, V. V. Flambaum, and A. Derevianko, Phys. Rev. Lett. **108**, 120802 (2012).

[151] U. I. Safronova, W. R. Johnson, and M. S. Safronova, Phys. Rev. A **76**, 042504 (2007).

[152] N. Herschbach, K. Pyka, J. Keller, and T. E. Mehlstäubler, Appl. Phys. B **107**, 891 (2012).

[153] G. A. Kazakov, A. N. Litvinov, V. I. Romanenko, L. P. Yatsenko, A. V Romanenko, M. Schreitl, G. Winkler, and T. Schumm, New J. Phys. **14**, 083019 (2012).

[154] D. A. Shirley, Rev. Mod. Phys. **36**, 339 (1964).

[155] C. Sanner, N. Huntemann, R. Lange, C. Tamm, E. Peik, M. S. Safronova, and S. G. Porsev, Nature **567**, 204 (2019).

[156] J.-P. Uzan, Living Rev. Relativ. **14**, 2 (2011).

[157] J.-P. Uzan, Comptes Rendus Phys. **16**, 576 (2015).

[158] M. S. Safronova, D. Budker, D. DeMille, D. F. J. Kimball, A. Derevianko, and C. W. Clark, Rev. Mod. Phys. **90**, 025008 (2018).

[159] J. C. Berengut, V. A. Dzuba, V. V. Flambaum, and S. G. Porsev, Phys. Rev. Lett. **102**, 210801 (2009).

[160] A. C. Hayes and J. L. Friar, Phys. Lett. B **650**, 229 (2007).

[161] P. Fadeev, J. C. Berengut, and V. V. Flambaum, Sensitivity of $^{229}$Th Nuclear Clock Transition to Variation of the Fine-Structure Constant, 2020.

[162] T. H. Dinh, A. Dunning, V. A. Dzuba, and V. V. Flambaum, Phys. Rev. A **79**, 054102 (2009).

[163] J. Guéna, M. Abgrall, D. Rovera, P. Rosenbusch, M. E. Tobar, P. Laurent, A. Clairon, and S. Bize, Phys. Rev. Lett. **109**, 080801 (2012).

[164] A. Derevianko and M. Pospelov, Nat. Phys. **10**, 933 (2014).

[165] A. Arvanitaki, J. Huang, and K. Van Tilburg, Phys. Rev. D - Part. Fields, Gravit. Cosmol. **91**, (2015).

[166] Y. V. Stadnik and V. V. Flambaum, Phys. Rev. Lett. **114**, (2015).

[167] G. Bertone and D. Hooper, Rev. Mod. Phys. **90**, (2018).

[168] F. Zwicky, Gen. Relativ. Gravit. **41**, (2009).

[169] R. Adam, P. A. R. Ade, N. Aghanim, Y. Akrami, M. I. R. Alves, F. Argüeso, M. Arnaud, F. Arroja, M. Ashdown, J. Aumont, C. Baccigalupi, M. Ballardini, A. J. Banday, R. B. Barreiro, J. G. Bartlett, N. Bartolo, S. Basak, P. Battaglia, E. Battaner, R. Battye, K. Benabed, A. Benoît, A. Benoit-Lévy, J.-P. Bernard, M. Bersanelli, B. Bertincourt, P. Bielewicz, I. Bikmaev, J. J. Bock, H. Böhringer, A. Bonaldi, L. Bonavera, J. R. Bond, J. Borrill, F. R. Bouchet, F. Boulanger, M. Bucher, R. Burenin, C. Burigana, R. C. Butler, E. Calabrese, J.-F. Cardoso, P. Carvalho, B. Casaponsa, G. Castex, A. Catalano, A. Challinor, A. Chamballu, R.-R. Chary, H. C. Chiang, J. Chluba, G. Chon, P. R. Christensen, S. Church, M. Clemens, D. L. Clements, S. Colombi, L. P. L. Colombo, C. Combet, B. Comis, D. Contreras, F. Couchot, A. Coulais, B. P. Crill, M. Cruz, A. Curto, F. Cuttaia, L. Danese, R. D. Davies, R. J. Davis, P. de Bernardis, A. de Rosa, G. de Zotti, J. Delabrouille, J.-M. Delouis, F.-X. Désert, E. Di Valentino, C. Dickinson, J. M. Diego, K. Dolag, H. Dole, S. Donzelli, O. Doré, M. Douspis, A. Ducout, J. Dunkley, X. Dupac, G. Efstathiou, P. R. M. Eisenhardt, F. Elsner, T. A. Enßlin, H. K. Eriksen, E. Falgarone, Y. Fantaye, M. Farhang, S. Feeney, J. Fergusson, R. Fernandez-Cobos, F. Feroz, F. Finelli, E. Florido, O. Forni, M. Frailis, A. A. Fraisse, C. Franceschet, E. Franceschi, A. Frejsel, A. Frolov, S. Galeotta, S. Galli, K. Ganga, C. Gauthier, R. T. Génova-Santos, M. Gerbino, T. Ghosh, M. Giard, Y. Giraud-Héraud, E. Giusarma, E. Gjerløw, J. González-Nuevo, K. M. Górski, K. J. B. Grainge, S. Gratton, A. Gregorio, A. Gruppuso, J. E. Gudmundsson, J. Hamann, W. Handley, F. K. Hansen, D. Hanson, D. L. Harrison, A. Heavens, G. Helou, S. Henrot-Versillé, C. Hernández-Monteagudo, D. Herranz, S. R. Hildebrandt, E. Hivon, M. Hobson, W. A. Holmes, A. Hornstrup, W. Hovest, Z. Huang, K. M. Huffenberger, G. Hurier, S. Ilić, A. H. Jaffe, T. R. Jaffe, T. Jin, W. C. Jones, M. Juvela, A. Karakci, E. Keihänen, R. Keskitalo, I. Khamitov, K. Kiiveri, J. Kim, T. S. Kisner, R. Kneissl, J. Knoche, L. Knox, N. Krachmalnicoff, M. Kunz, H. Kurki-Suonio, F. Lacasa, G. Lagache, A. Lähteenmäki, J.-M. Lamarre, M. Langer, A. Lasenby, M. Lattanzi, C. R. Lawrence, M. Le Jeune, J. P. Leahy, E. Lellouch, R. Leonardi, J. León-Tavares, J. Lesgourgues, F. Levrier, A. Lewis, M. Liguori, P. B. Lilje, M. Lilley, M. Linden-Vørnle, V. Lindholm, H. Liu, M. López-Caniego, P. M. Lubin, Y.-Z. Ma, J. F. Macías-Pérez, G. Maggio, D. Maino, D. S. Y. Mak, N. Mandolesi, A. Mangilli, A. Marchini, A. Marcos-Caballero, D. Marinucci, M. Maris, D. J. Marshall, P. G.

Martin, M. Martinelli, E. Martínez-González, S. Masi, S. Matarrese, P. Mazzotta, J. D. McEwen, P. McGehee, S. Mei, P. R. Meinhold, A. Melchiorri, J.-B. Melin, L. Mendes, A. Mennella, M. Migliaccio, K. Mikkelsen, M. Millea, S. Mitra, M.-A. Miville-Deschênes, D. Molinari, A. Moneti, L. Montier, R. Moreno, G. Morgante, D. Mortlock, A. Moss, S. Mottet, M. Münchmeyer, D. Munshi, J. A. Murphy, A. Narimani, P. Naselsky, A. Nastasi, F. Nati, P. Natoli, M. Negrello, C. B. Netterfield, H. U. Nørgaard-Nielsen, F. Noviello, D. Novikov, I. Novikov, M. Olamaie, N. Oppermann, E. Orlando, C. A. Oxborrow, F. Paci, L. Pagano, F. Pajot, R. Paladini, S. Pandolfi, D. Paoletti, B. Partridge, F. Pasian, G. Patanchon, T. J. Pearson, M. Peel, H. V. Peiris, V.-M. Pelkonen, O. Perdereau, L. Perotto, Y. C. Perrott, F. Perrotta, V. Pettorino, F. Piacentini, M. Piat, E. Pierpaoli, D. Pietrobon, S. Plaszczynski, D. Pogosyan, E. Pointecouteau, G. Polenta, L. Popa, G. W. Pratt, G. Prézeau, S. Prunet, J.-L. Puget, J. P. Rachen, B. Racine, W. T. Reach, R. Rebolo, M. Reinecke, M. Remazeilles, C. Renault, A. Renzi, I. Ristorcelli, G. Rocha, M. Roman, E. Romelli, C. Rosset, M. Rossetti, A. Rotti, G. Roudier, B. Rouillé d'Orfeuil, M. Rowan-Robinson, J. A. Rubiño-Martín, B. Ruiz-Granados, C. Rumsey, B. Rusholme, N. Said, V. Salvatelli, V. Salvati, M. Sandri, H. S. Sanghera, D. Santos, R. D. E. Saunders, A. Sauvé, M. Savelainen, G. Savini, B. M. Schaefer, M. P. Schammel, D. Scott, M. D. Seiffert, P. Serra, E. P. S. Shellard, T. W. Shimwell, M. Shiraishi, K. Smith, T. Souradeep, L. D. Spencer, M. Spinelli, S. A. Stanford, D. Stern, V. Stolyarov, R. Stompor, A. W. Strong, R. Sudiwala, R. Sunyaev, P. Sutter, D. Sutton, A.-S. Suur-Uski, J.-F. Sygnet, J. A. Tauber, D. Tavagnacco, L. Terenzi, D. Texier, L. Toffolatti, M. Tomasi, M. Tornikoski, D. Tramonte, M. Tristram, A. Troja, T. Trombetti, M. Tucci, J. Tuovinen, M. Türler, G. Umana, L. Valenziano, J. Valiviita, F. Van Tent, T. Vassallo, L. Vibert, M. Vidal, M. Viel, P. Vielva, F. Villa, L. A. Wade, B. Walter, B. D. Wandelt, R. Watson, I. K. Wehus, N. Welikala, J. Weller, M. White, S. D. M. White, A. Wilkinson, D. Yvon, A. Zacchei, J. P. Zibin, and A. Zonca, Astron. Astrophys. **594**, A1 (2016).

[170] L. Posti and A. Helmi, Astron. Astrophys. **621**, (2019).
[171] S. L. Adler, Solar System Dark Matter, 2009.
[172] P. S. Corasaniti, S. Agarwal, D. J. E. Marsh, and S. Das, Phys. Rev. D **95**, (2017).
[173] A. Derevianko, Phys. Rev. A **97**, (2018).
[174] K. Van Tilburg, N. Leefer, L. Bougas, and D. Budker, Phys. Rev. Lett. **115**, (2015).
[175] A. Hees, J. Guéna, M. Abgrall, S. Bize, and P. Wolf, Phys. Rev. Lett. **117**, (2016).
[176] D. Antypas, O. Tretiak, A. Garcon, R. Ozeri, G. Perez, and D. Budker, Phys. Rev. Lett. **123**, (2019).
[177] C. J. Kennedy, E. Oelker, J. M. Robinson, T. Bothwell, D. Kedar, W. R. Milner, G. E. Marti, A. Derevianko, and J. Ye, (2020).
[178] B. M. Roberts, G. Blewitt, C. Dailey, M. Murphy, M. Pospelov, A. Rollings, J. Sherman, W. Williams, and A. Derevianko, Nat. Commun. **8**, 1195 (2017).
[179] P. Wcisło, P. Morzyński, M. Bober, A. Cygan, D. Lisak, R. Ciuryło, and M. Zawada, Nat. Astron. **1**, (2017).
[180] P. Wcisło, P. Ablewski, K. Beloy, S. Bilicki, M. Bober, R. Brown, R. Fasano, R. Ciuryło, H. Hachisu, T. Ido, J. Lodewyck, A. Ludlow, W. McGrew, P. Morzyński, D. Nicolodi, M. Schioppa, M. Sekido, R. Le Targat, P. Wolf, X. Zhang, B. Zjawin, and M. Zawada, Sci. Adv. **4**, (2018).
[181] B. M. Roberts, P. Delva, A. Al-Masoudi, A. Amy-Klein, C. Bærentsen, C. F. A. Baynham, E. Benkler, S. Bilicki, S. Bize, W. Bowden, J. Calvert, V. Cambier, E. Cantin, E. A. Curtis, S. Dörscher, M. Favier, F. Frank, P. Gill, R. M. Godun, G. Grosche, C. Guo, A. Hees, I. R. Hill, R. Hobson, N. Huntemann, J. Kronjäger, S. Koke, A. Kuhl, R. Lange, T. Legero, B. Lipphardt, C. Lisdat, J. Lodewyck, O. Lopez, H. S. Margolis, H. Álvarez-Martínez, F. Meynadier, F. Ozimek, E. Peik, P.-E. Pottie, N. Quintin, C. Sanner, L. De Sarlo, M. Schioppo, R. Schwarz, A. Silva, U. Sterr, C. Tamm, R. Le Targat, P. Tuckey, G. Vallet, T. Waterholter, D. Xu, and P. Wolf, New J. Phys. **22**, (2020).
[182] A. Borrelli and E. Castellani, Found. Phys. **49**, (2019).
[183] A. Merle and T. Ohlsson, Nat. Phys. **8**, 584 (2012).
[184] P. W. Graham, D. E. Kaplan, and S. Rajendran, Phys. Rev. Lett. **115**, (2015).
[185] J. R. Espinosa, C. Grojean, G. Panico, A. Pomarol, O. Pujolàs, and G. Servant, Phys. Rev. Lett. **115**, (2015).
[186] A. Banerjee, H. Kim, and G. Perez, Phys. Rev. D **100**, (2019).
[187] S. A. Abel, R. S. Gupta, and J. Scholtz, Phys. Rev. D **100**, (2019).
[188] A. Banerjee, H. Kim, O. Matsedonskyi, G. Perez, and M. S. Safronova, J. High Energy Phys. **2020**, (2020).